\documentclass[bibyear]{aa}
\usepackage[varg]{txfonts}
\usepackage{graphicx}

\def\I{\mbox{IRAS\,20126+4104}}

\def\Msun{\mbox{$M_\odot$}}
\def\Lsun{\mbox{$L_\odot$}}
\def\WAT{H$_2$O}

\def\COI{\mbox{$^{13}$CO}}
\def\COII{\mbox{C$^{18}$O}}

\def\FORM{\mbox{H$_2$CO}}
\def\METH{\mbox{CH$_3$OH}}
\def\MCN{\mbox{CH$_3$CN}}
\def\MCNI{\mbox{CH$_3^{13}$CN}}
\def\MCNII{\mbox{$^{13}$CH$_3$CN}}

\def\HM{\mbox{H$_2$}}
\def\HII{H{\sc ii}}

\def\kms{\mbox{km~s$^{-1}$}}
\def\cmc{cm$^{-3}$}

\def\mic{\mbox{$\mu$m}}

\def\Eup{\mbox{$E_{\rm u}$}}
\def\d{{\rm d}}
\def\e{{\rm e}}

\def\Tb{\mbox{$T_{\rm B}$}}
\def\Tk{\mbox{$T_{\rm K}$}}
\def\Trot{\mbox{$T_{\rm rot}$}}

\def\Ncol{\mbox{$N_{\rm col}$}}
\def\xs{\mbox{$X_{\rm s}$}}
\def\ys{\mbox{$Y_{\rm s}$}}
\def\Vs{\mbox{$V_{\rm sys}$}}
\def\Ms{\mbox{$M_{\rm s}$}}
\def\PA{{\rm PA_d}}
\def\PAd{{\rm PA_{da}}}
\def\PAj{{\rm PA_{ja}}}
\def\Rd{R_{\rm d}}

\begin{document}

\title{
Dissecting the disk and the jet of a massive (proto)star
}
\subtitle{
An ALMA view of \I}
\author{
        R.~Cesaroni\inst{1} \and\
        D.~Galli\inst{1} \and\ M. Padovani\inst{1}
        \and V.~M.~Rivilla\inst{2} \and\ \'A.~S\'anchez-Monge\inst{3,4}
}
\institute{
 INAF, Osservatorio Astrofisico di Arcetri, Largo E. Fermi 5, I-50125 Firenze, Italy
           \email{riccardo.cesaroni@inaf.it}
 \and
 Centro de Astrobiolog\'{\i}a (CAB), CSIC-INTA, Carretera de Ajalvir km 4, Torrej\'on de Ardoz, E-28850 Madrid, Spain
 \and
 Institut de Ci\`encies de l'Espai (ICE, CSIS), Carrer de Can Magrans s/n, E-08193 Bellaterra, Spain
 \and
 Institut d'Estudis Espacials de Catalunya (IEEC), Barcelona, Spain
}
\offprints{R. Cesaroni, \email{riccardo.cesaroni@inaf.it}}
\date{Received date / Accepted date}

\abstract 
{
The study of disks around early-type (proto)stars has recently been boosted
by a new generation of instruments, and additional evidence has been found of disk+jet
systems around stars of up to $\sim$20~\Msun\ . These results
appear to confirm theoretical predictions that even the most massive stars
may form though disk-mediated accretion.
}{
We want to investigate one of the best examples of disk+jet systems around
an early B-type (proto)star, \I. The relatively simple structure of this
object and its relative proximity to Earth (1.64~kpc) make it an ideal target for
resolution of its disk and the determination of its physical and kinematical structure.
}{
Despite the high declination of \I, it has been possible to perform successful
observations with the Atacama Large Millimeter and submillimeter Array
at 1.4~mm in the continuum emission and a number of molecular tracers of
high-density gas (for the disk) and shocked gas (for the jet).
}{
The new data allow us to improve on previous similar observations of \I\ and
confirm the existence of a Keplerian accretion disk around a $\sim$12~\Msun\
(proto)star. From methyl cyanide, we derived the rotation
temperature and column density as a function of disk radius. We also obtained a map
of the same quantities for the jet using the ratio
between two lines of formaldehyde. We also use two simple models of
the jet and the disk to estimate the basic geometrical and kinematical
parameters of the two. From the temperature and column density profiles,
we conclude that the disk is stable at all radii. We also estimate an accretion
rate of $\sim$$10^{-3}$~\Msun~yr$^{-1}$.
}{
Our analysis confirms that the jet from \I\ is highly collimated, lies
close to the plane of the sky, and expands with velocity increasing with
distance. As expected, the gas temperature and column density peak in the
bow shock. The disk is undergoing Keplerian rotation but a non-negligible
radial velocity component is also present that is equal to $\sim$40\% of the
rotational component. The disk is slightly inclined with respect to the
line of sight and has a dusty envelope that absorbs
the emission from the disk surface. This causes a slight distortion of
the disk structure
observed in high-density tracers such as methyl cyanide. We also reveal
a significant deviation from axial symmetry in the SW part of the disk,
which might be caused by either tidal interaction with a nearby, lower-mass
companion or interaction with the outflowing gas of the jet.
}

\maketitle

\section{Introduction}
\label{sint}

The important role played by disks in the process of star formation is
amply evidenced by the plenitude of proto-planetary disks found around
Solar-type (proto)stars in recent years (e.g., Benisty et al.~\cite{beni}).
The impressive images obtained with the Atacama Large Millimeter and
submillimeter Array (ALMA) have allowed extremely detailed studies of the disk structure and have revealed the presence of a variety of structures from gaps
to spiral arms (e.g., Garufi et al.~\cite{garu}, Andrews~\cite{andr}). The
situation is quite different for early-type, and especially O-type,
(proto)stars. While ALMA observations have
increased the number of candidate circumstellar disks around massive
(proto)stars, none of these resemble the disks around their low-mass
siblings. A reason for this difference is certainly the large distances
to these objects (typically a few kiloparsecs), which imply linear resolutions of
$>$100~au. Another reason is that high-mass stars are born deeply embedded
in their parental, optically thick cores, which makes it difficult to
distinguish the disk from the envelope enshrouding it. In addition, the
presence of rich stellar clusters, besides complicating the observations,
may also affect the disk structure by tidal interactions, thus altering
the disk morphology and velocity field; and multiple (proto)stars
power multiple outflows, which makes the picture even more confused.

For these reasons, we must focus on the
most promising disk candidates around massive stars, as this strategy allows us to obtain the
greatest amount of information on these objects through high-resolution observations.
One of the first and best studied disk+jet systems associated with a massive
young stellar object (YSO) is certainly \I. This object has been observed in a
large number of tracers over a broad range of wavelengths, spanning from
the centimeter regime to X-rays. Here, we summarize the main findings from the literature that
are the most relevant to the present study and explain why we decided to focus our
attention on \I\ in  particular.

The distance estimate, obtained from parallax measurements of the \WAT\
masers, ranges from 1.33$^{+0.19}_{-0.15}$ (Hirota et al.~\cite{vera})
to 1.64$\pm$0.05~kpc (Moscadelli et al.~\cite{mosca11}), which are consistent with
each other to within 1.3~$\sigma$. In the following, we adopt a distance of 1.64~kpc
because of the smaller error and for consistency with previous
studies of \I. The source appears to be relatively isolated in the near-infrared
(Qiu et al.~\cite{qiu08}), but X-ray observations have revealed a deeply
embedded cluster associated with it (Montes et al.~\cite{mont15}). The main
feature of \I\ is the presence of a circumstellar structure, which appears
to undergo Keplerian rotation around a (proto)star of $\sim$$10^4$~\Lsun\
(Cesaroni et al.~\cite{cesa97,cesa99,cesa23}). Model fits to the spectral
energy distribution as well as to the molecular line data have been
performed by several authors (Keto \& Zhang~\cite{ketzha}; Johnston et
al.~\cite{johns}; Chen et al.~\cite{chen}), who derive a stellar mass
of $\sim$12~\Msun. The lack of a detectable \HII\ region around such a
massive and luminous star suggests that accretion might still be present, quenching the formation of such a region (see Walmsley~\cite{walms}). The object
is also powering a thermal radio jet (Hofner et al.~\cite{hofn99,hofn07}),
and exhibits a molecular outflow (Wilking et al.~\cite{wilk90}; Cesaroni et
al.~\cite{cesa97,cesa99}; Kawamura et al.~\cite{kawa99}) that is roughly
perpendicular to the disk and is undergoing precession, probably due to a
nearby star (Shepherd et al.~\cite{shep}; Cesaroni et al.~\cite{cesa05};
Caratti o Garatti et al.~\cite{caga08}). The expansion of the jet has been
measured by means of the \WAT\ maser proper motions down to $\sim$100~au
from the star (Moscadelli et al.~\cite{mosca05,mosca11}) and with multi-epoch
imaging of the \HM\ jet up to $\sim$$10^4$~au (Massi et al.~\cite{massi23}).

Given the relative proximity and simple structure of the source,
which make \I\ an ideal disk+jet system for study, we
decided to use the most powerful millimeter interferometer available
to date in order to analyze the structure and kinematics of this object. Despite
the high declination of the source, the observations are feasible with ALMA,
and the new data are an improvement by at least a factor of $\sim$2 in angular
resolution and $\sim$5 in sensitivity with respect to previous observations
at the same wavelengths (Cesaroni et al.~\cite{cesa99,cesa14}; Su et
al.~\cite{su07}).

\section{Observations and data reduction}
\label{sobs}

The ALMA observations of \I\ were performed in band~6 at three epochs, on
May 20, 2017, July 6, 2017, and October 25, 2021 (projects 2016.1.00595.S and
2019.1.00199.S).
The phase center of the observations was
$\alpha$(ICRS)=$20^{\rm h}14^{\rm m}26\fs0364$
and $\delta$(ICRS)=41\degr13\arcmin32\farcs516.
For the observations made in 2017, J2025+3343 and J2148+0657 were used as bandpass and flux calibrator,
respectively, whereas in 2021, J2253+1608
was used for both calibrations. In all cases, the phase calibrator was
J2015+3710. The correlator configuration consisted of one unit of 1.875~GHz
and 12 units of 234~MHz.

The data were calibrated with the ALMA data-reduction pipeline and
subsequent data flagging and imaging were made with the Common Astronomy
Software Applications (CASA) software\footnote{CASA can be downloaded
from http://casa.nrao.edu}. We created cubes with natural weighting
covering all of the correlator units, using only the data obtained
with the compact configuration in 2017. The data were corrected for
primary beam attenuation. The continuum was subtracted with the STATCONT
software\footnote{https://hera.ph1.uni-koeln.de/$\sim$sanchez/statcont}
developed by S\'anchez-Monge et al.~(\cite{statcont}). Then all line cubes
were inspected to search for possible outflow tracers besides the $J$=2--1
transition of the well-known CO isotopologs and the SiO(5--4) line.
As a result, we found a further four lines that originate from the outflow
lobes, namely the \FORM($3_{0,3}$--$2_{0,2}$), ($3_{2,2}$--$2_{2,1}$), and
($3_{2,1}$--$2_{2,0}$) transitions and the \METH($4_{-2,3}$--$3_{-1,2}$)~E line.
In practice, only the SiO, \FORM, and \METH\ data were usable for our study,
because the \COI\ and \COII\ line
emission
were heavily resolved out.
The synthesized beam of the images is about $0\farcs48\times0\farcs28$ with
a position angle (PA) of -7\degr.

We also created cubes with natural weighting using all the available
configurations (i.e., the data from both 2017 and 2021) only for the
correlator unit covering the largest bandwidth (1.875~GHz) and those
containing selected molecular species that we wish to use for studying
the disk, specifically the \MCN\ isotopologs and DCN. We then obtained
a continuum image from the cube spanning the 1.875~GHz bandwidth with
\mbox{STATCONT}, whereas we preferred to subtract the continuum from the
other cubes by removing a constant baseline fitted to line-free
channels selected by visual inspection of the spectra. The resulting synthesized
beam is approximately $0\farcs23\times0\farcs13$ with a PA of --12\degr.
The final images were converted to GILDAS format and analyzed with the
GILDAS\footnote{The GILDAS software has been developed at IRAM and
Observatoire de Grenoble abd is available at http://iram.fr/IRAMFR/GILDAS/}
software.

\begin{figure}
\centering
\resizebox{8.5cm}{!}{\includegraphics[angle=0]{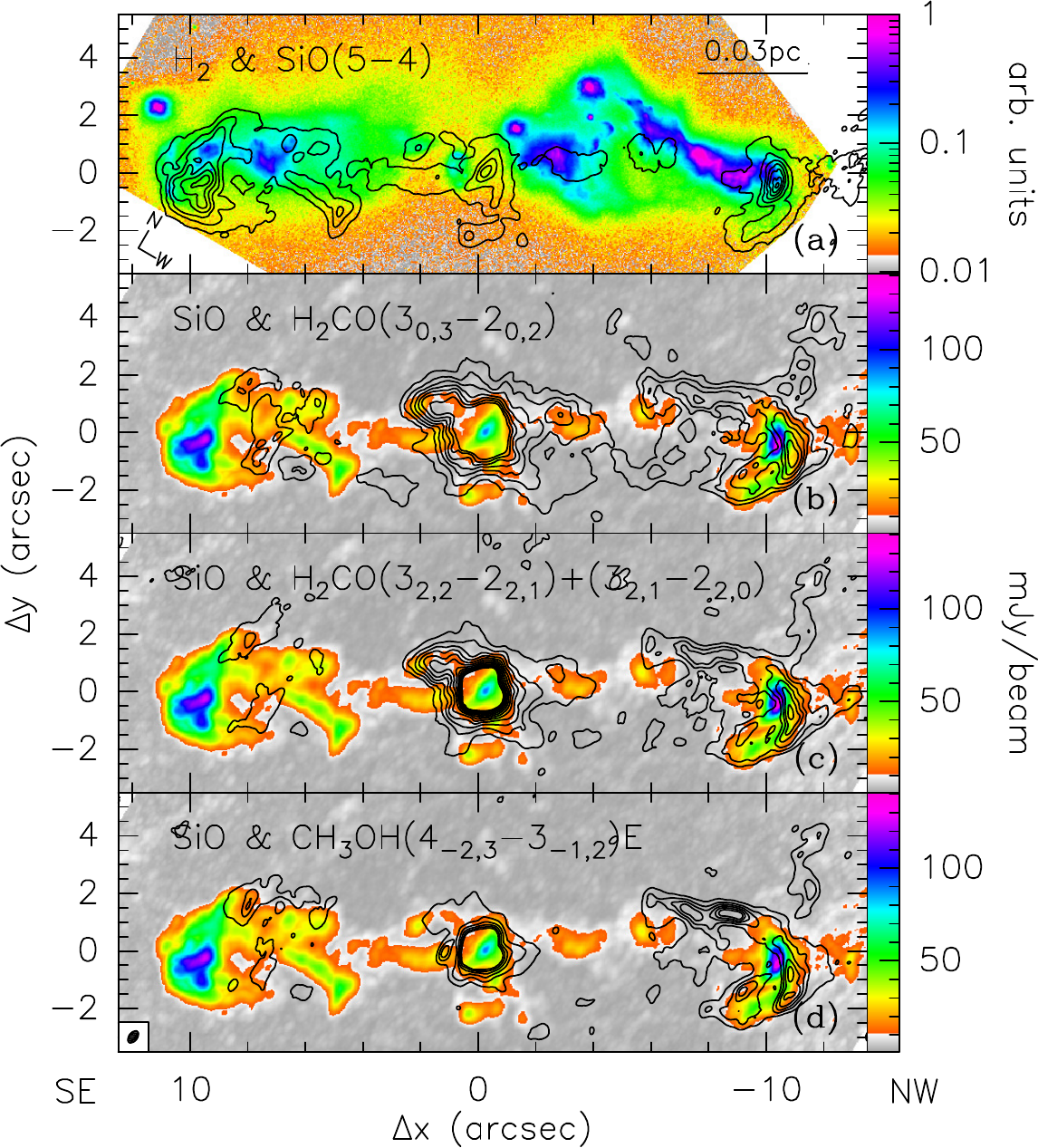}}
\caption{
Moment-8 maps in various molecular tracers of the jet in \I.  All maps have
been rotated by 30\degr\ clockwise so that the jet axis lies parallel to
the X-axis of the plot. The offset is relative to the phase center of the
observations. The ellipse in the bottom left corner indicates
the synthesized beam of the ALMA maps.
{\bf (a)} Contour map of the SiO(5--4) line emission overlaid on the
image of the \HM\ 2.12~\mic\ line obtained by
Cesaroni et al.~(\cite{cesa13}). Contour levels range from 10.4 to
136.4 in steps of 21 mJy/beam.
{\bf (b)} Contour map of the \FORM($3_{0,3}$--$2_{0,2}$) line emission
overlaid on the same SiO(5--4) image as in panel~(a). Contour levels range
from 27 to 117 in steps of 15~mJy/beam.
{\bf (c)} Same as panel (b) but for the contour map of the average emission
in the \FORM($3_{2,2}$--$2_{2,1}$) and ($3_{2,1}$--$2_{2,0}$) lines.
Contour levels range from 12 to 111 in steps of 9 mJy/beam.
{\bf (d)} Same as panel (b) but for the contour map of the
\METH($4_{-2,3}$--$3_{-1,2}$)~E line emission. Contour levels range from
16.2 to 111.2 in steps of 19 mJy/beam.
}
\label{fjmaps}
\end{figure}

\section{Results}
\label{sres}

Based on previous studies (see, e.g., Cesaroni et al.~\cite{cesa99,cesa14}
and Su et al.~\cite{su07}), we know that the disk in \I\ is best imaged in the
millimeter continuum emission and in complex organic molecules (COMs)
such as \MCN, whereas a typical shock tracer observed in the outflowing
gas is SiO. For the sake of simplicity, hereafter we refer to the
collimated bipolar structure observed in SiO as the ``jet''. This structure
is  also detected in other molecules (see Sect.~\ref{srjet})
prominent in the disk. In this section, we describe the results obtained
for the disk and the jet from the analysis of selected molecular species,
namely three \MCN\ isotopologs, DCN, \METH, \FORM, and SiO.

\subsection{The jet}
\label{srjet}

\begin{figure}
\centering
\resizebox{8.5cm}{!}{\includegraphics[angle=0]{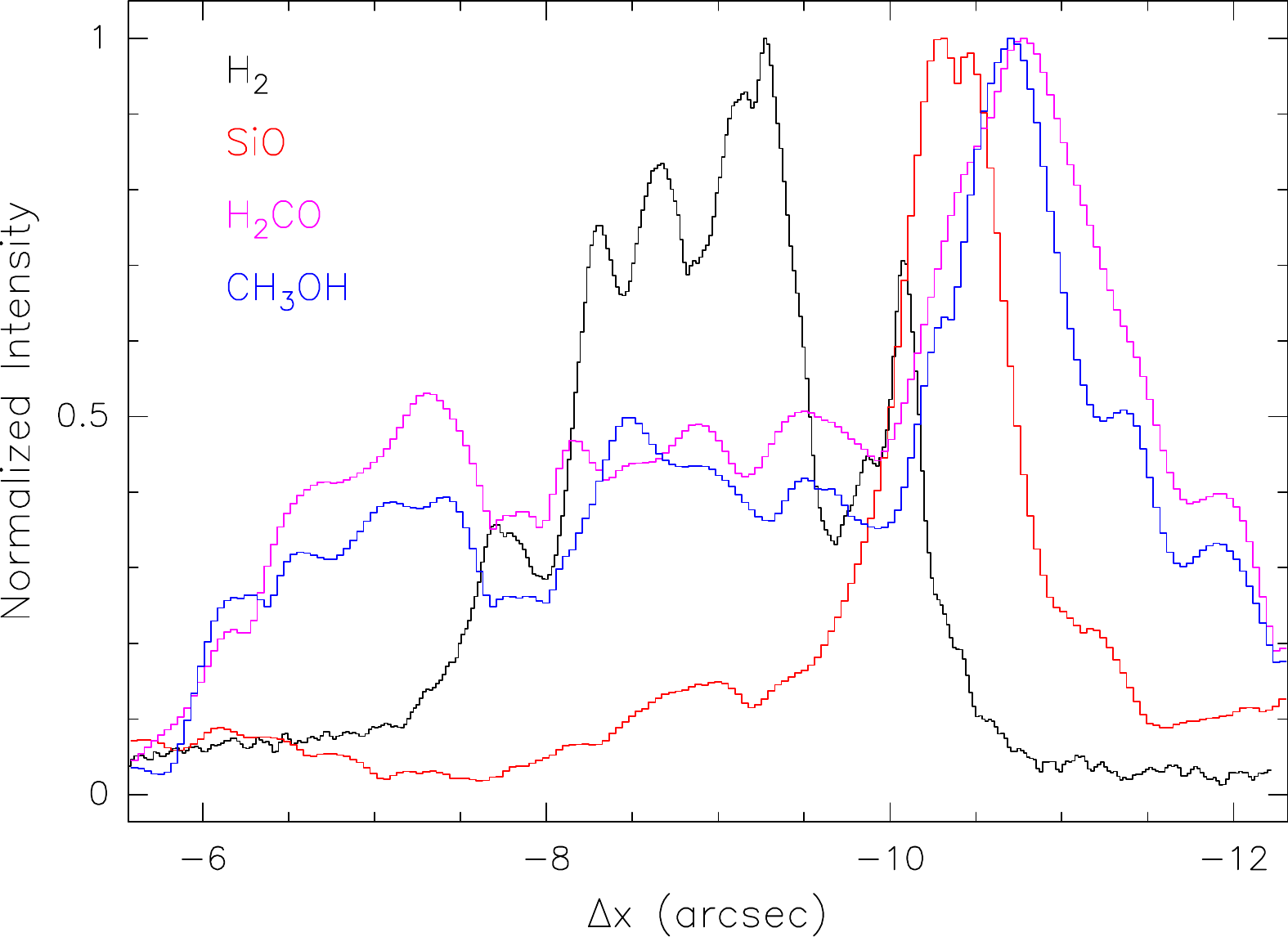}}
\caption{
Normalized intensity profile along the jet axis. The X-axis is the same
as in Fig.~\ref{fjmaps} but spans only the range of angles corresponding
to the NW lobe of the jet. The meaning of the different curves is given
in the figure.
}
\label{fcut}
\end{figure}

As mentioned in Sect.~\ref{sint}, the jet+outflow from \I\ has been
studied in detail by several authors. Here we take advantage of the high
sensitivity of our ALMA observations to perform an improved analysis of the
jet structure and physical parameters on scales of up to $\sim$0.06~pc or
12000~au. For this purpose, we inspected the data cubes covering
the whole frequency range of our correlator setup and found five lines
belonging to three molecular species that appear to trace the jet, namely the
SiO(5--4) (upper level energy \Eup=31~K), \METH($4_{-2,3}$--$3_{-1,2}$)~E
(\Eup=45~K), and \FORM($3_{0,3}$--$2_{0,2}$) (\Eup=21~K),
($3_{2,2}$--$2_{2,1}$) (\Eup=68~K) and ($3_{2,1}$--$2_{2,0}$) (\Eup=68~K)
transitions. As the latter two transitions have the same excitation
energies and line strengths, we decided to average the two data sets to
improve the signal-to-noise ratio (S/N). A comparison among the moment-8
maps\footnote{The so-called moment-8 maps are obtained by selecting the
maximum intensity across the spectrum for each pixel of the image. With
respect to the maps obtained by averaging the line emission over a broad
velocity range, the moment-8 maps have the advantage of also being sensitive
to faint, narrow lines.}
of all these lines and the \HM\ 2.12~\mic\ line imaged by Cesaroni
et al.~(\cite{cesa13}) and Massi et al.~(\cite{massi23}) is shown in
Fig.~\ref{fjmaps}. We note that, unlike the bow shock to the SE, which
is detected only in the SiO line, the one to the NW is traced by all
molecules. It is also interesting to note that the emission in different
molecules peaks at different places in the NW bow shock. In particular,
the \HM\ emission appears to lie downstream
(i.e., closer to the center of ejection)
with respect to the SiO shock,
whereas both the \METH\ and \FORM\ emissions are located upstream
(i.e., farther from the center of ejection),
an effect
that can be better appreciated in Fig.~\ref{fcut}, where we show the emission
in the different tracers along the jet axis passing through the NW lobe.
This configuration
could mirror
a chemical stratification possibly
due to temperature and density gradients across the bow shock,
although excitation effects due to the variation of these
physical parameters cannot be excluded.

\begin{figure}
\centering
\resizebox{8.5cm}{!}{\includegraphics[angle=0]{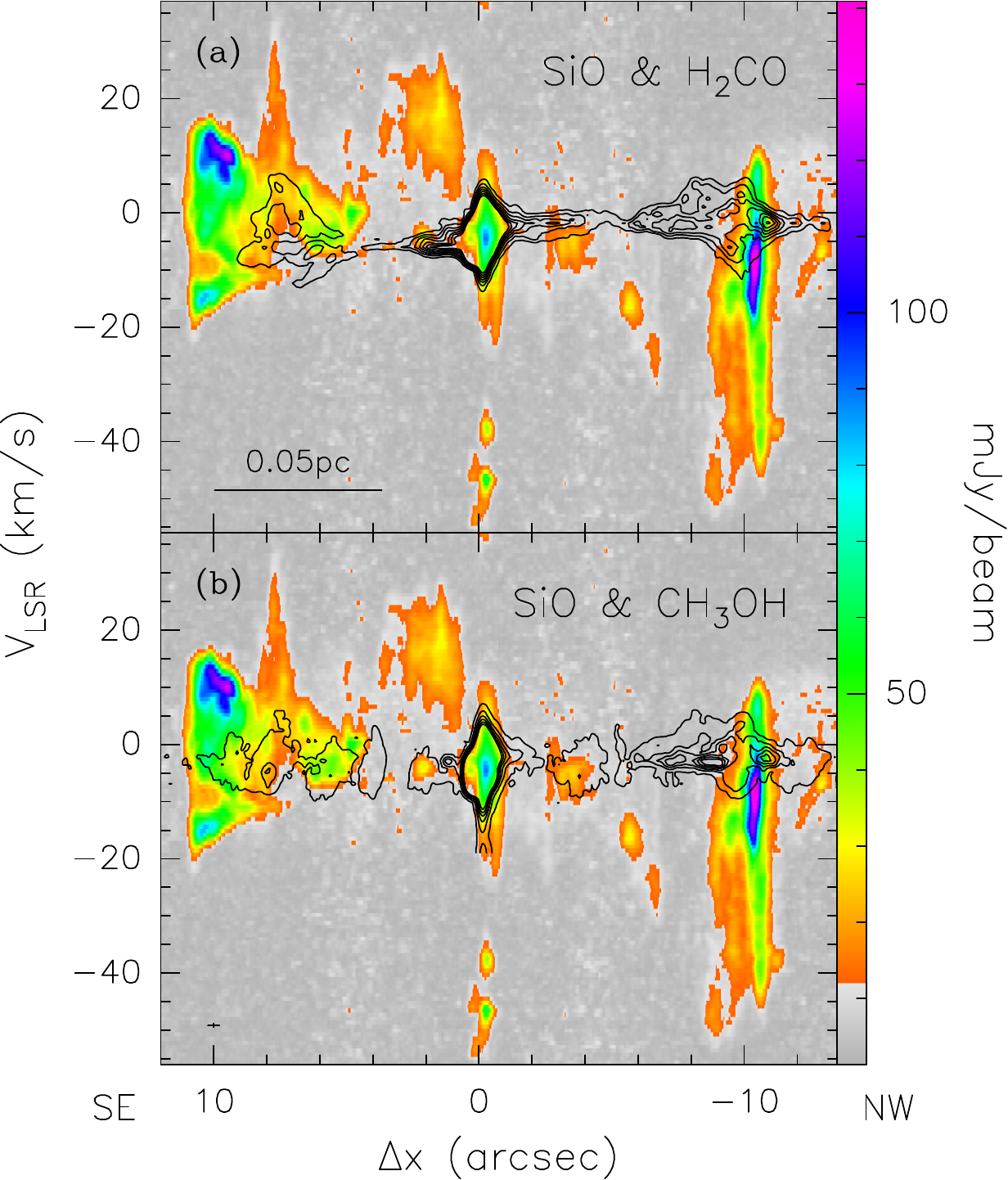}}
\caption{
Moment-8 PV diagrams in three molecular tracers of the jet along
the direction with a PA of --60\degr. The offset is relative to the
phase center of the observations.
The cross in the bottom left corner indicates the angular and spectral
resolutions.
{\bf (a)} Contour map of the \FORM($3_{0,3}$--$2_{0,2}$) line emission
overlaid on the SiO(5--4) map. Contour levels range
from 22 to 112 in steps of 15~mJy/beam.
{\bf (b)} Same as panel (a) but for the contour map of the
\METH($4_{-2,3}$--$3_{-1,2}$)~E line emission. Contour levels range from
16.2 to 111.2 in steps of 19 mJy/beam.
}
\label{fjpvs}
\end{figure}

The complementary distribution of the different tracers is observed
also in the velocity field of the bow shocks, which is revealed by the
position--velocity (PV) diagrams in Fig.~\ref{fjpvs} along the jet axis
(i.e., along a cut with $\PAj$$\simeq$$-60\degr$). These
were obtained with the same method used for the moment-8 maps:
for each pixel of the PV diagram, we took the maximum emission
along the direction perpendicular to the jet axis. Clearly, the \METH\
and \FORM\ emissions cover approximately the same area in the plot, and
both are complementary to the SiO emission. We note that the
contours of the PV diagrams have been chosen in such a way to emphasise
the emission from the lobes, while that from the core is illustrated in
Sect.~\ref{srdisk}. However, it is clear that the strongest \FORM\ and
\METH\ emission is seen around the star, unlike the SiO emission, which
is dominant at the tips of the jet lobes, as expected for a shock tracer.
We analyze the structure of the bow shock in more detail in Sect.~\ref{sajet}.

\subsection{The disk}
\label{srdisk}

\begin{figure}
\centering
\resizebox{8.5cm}{!}{\includegraphics[angle=0]{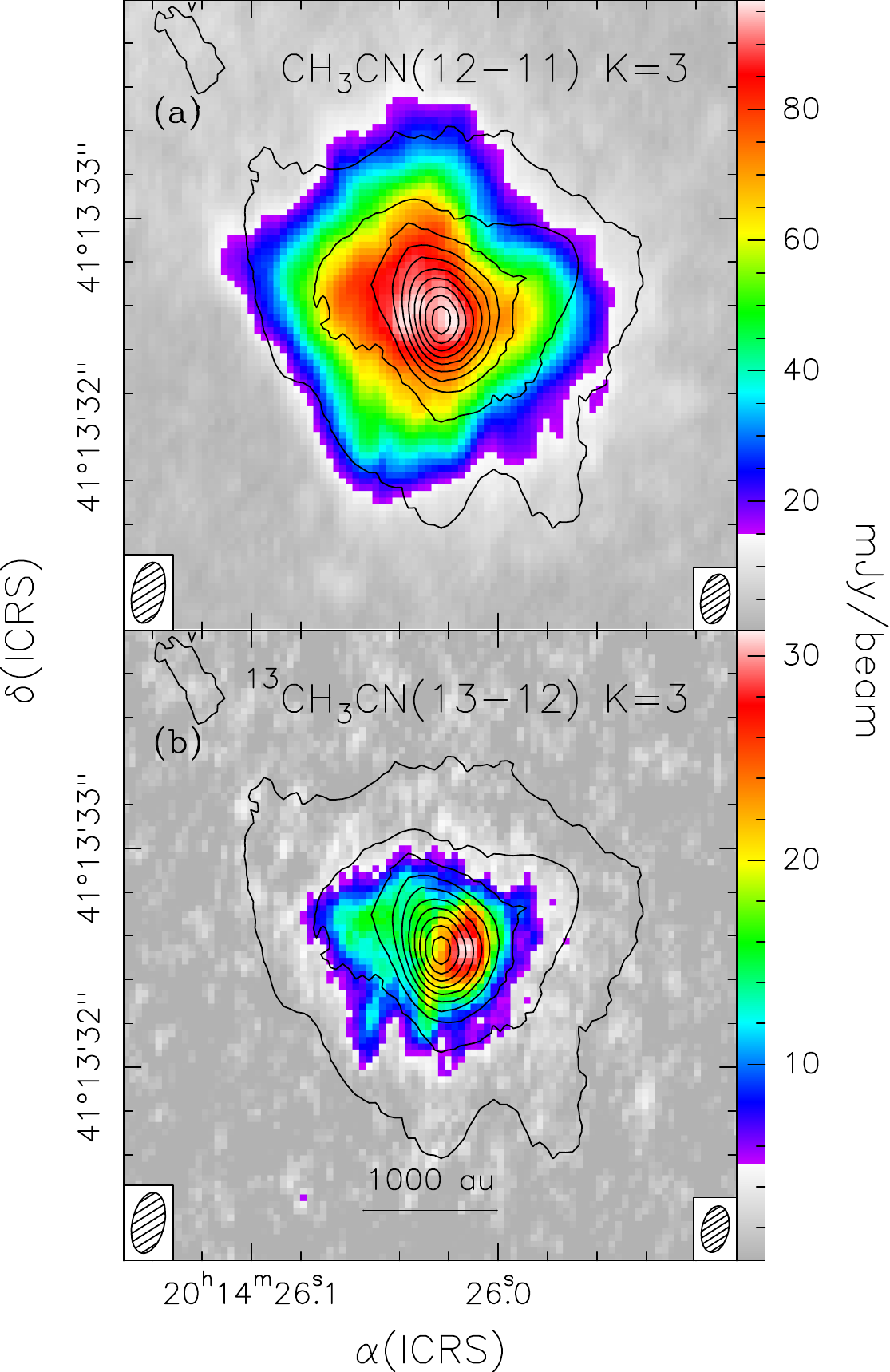}}
\caption{
Map of the 1.4~mm continuum emission (contours) overlaid on the moment-8
map of the \MCN(12--11) $K$=3 (panel {\bf a}) and \MCNII(13--12) $K$=3
line (panel {\bf b}). Contour levels range from 1.4 to 26.6 in steps
of 2.8~mJy/beam. The ellipses in the bottom left and right corners
indicate the synthesized beam of the continuum and line maps, respectively.
}
\label{fcoli}
\end{figure}

The contour map in Fig.~\ref{fcoli} shows that the continuum emission
is made of an extended halo with a barely resolved peak of emission at
the center, which in the following we assume to pin-point the
position of the (proto)star. This morphology suggests that the continuum is
arising from both the disk and a surrounding envelope.
We fitted the continuum map with a 2D Gaussian
using task {\em imfit} of CASA. The position of the 1.4~mm continuum peak
obtained in this way provides us with an estimate of
the disk center (i.e., the position of the star), which is
$\alpha$(J2000)=$20^{\rm h}14^{\rm m}26\fs022\pm0\fs0011$
and $\delta$(J2000)=41\degr13\arcmin32\farcs58$\pm$0\farcs013.
The deconvolved full width at half maximum (FWHM) is
(0\farcs71$\pm$0\farcs04)$\times$(0\farcs58$\pm$0\farcs04) or
(1160$\pm$70)$\times$(950$\pm$70)~au with a PA of 56\degr$\pm$15\degr. We
note that the latter value is consistent within the error with the PA of
the major axis of the disk derived by other authors (53\degr; Cesaroni et
al.~\cite{cesa05}) from their \MCN\ line data.

The emission from the disk is illustrated in Fig.~\ref{fcoli} by the moment-8
maps of the \MCN(12--11) and \MCNII(13--12) line emission. We have chosen
the $K$=3 component, which is the least affected by blending with other
transitions. As expected, the thinner \MCNII\ line emission is detected
over a more compact region than the main species, while the extension of
the latter approximately matches that of the continuum emission. Another
obvious difference between Fig.~\ref{fcoli}a and~\ref{fcoli}b is that in
the former the emission is maximum around the continuum peak, whereas in
the latter the peak of the emission is clearly shifted to the WSW. A
similar behavior is seen in the PV diagrams, as discussed later in
Sect.~\ref{skep}.

\begin{figure}
\centering
\resizebox{8.5cm}{!}{\includegraphics[angle=0]{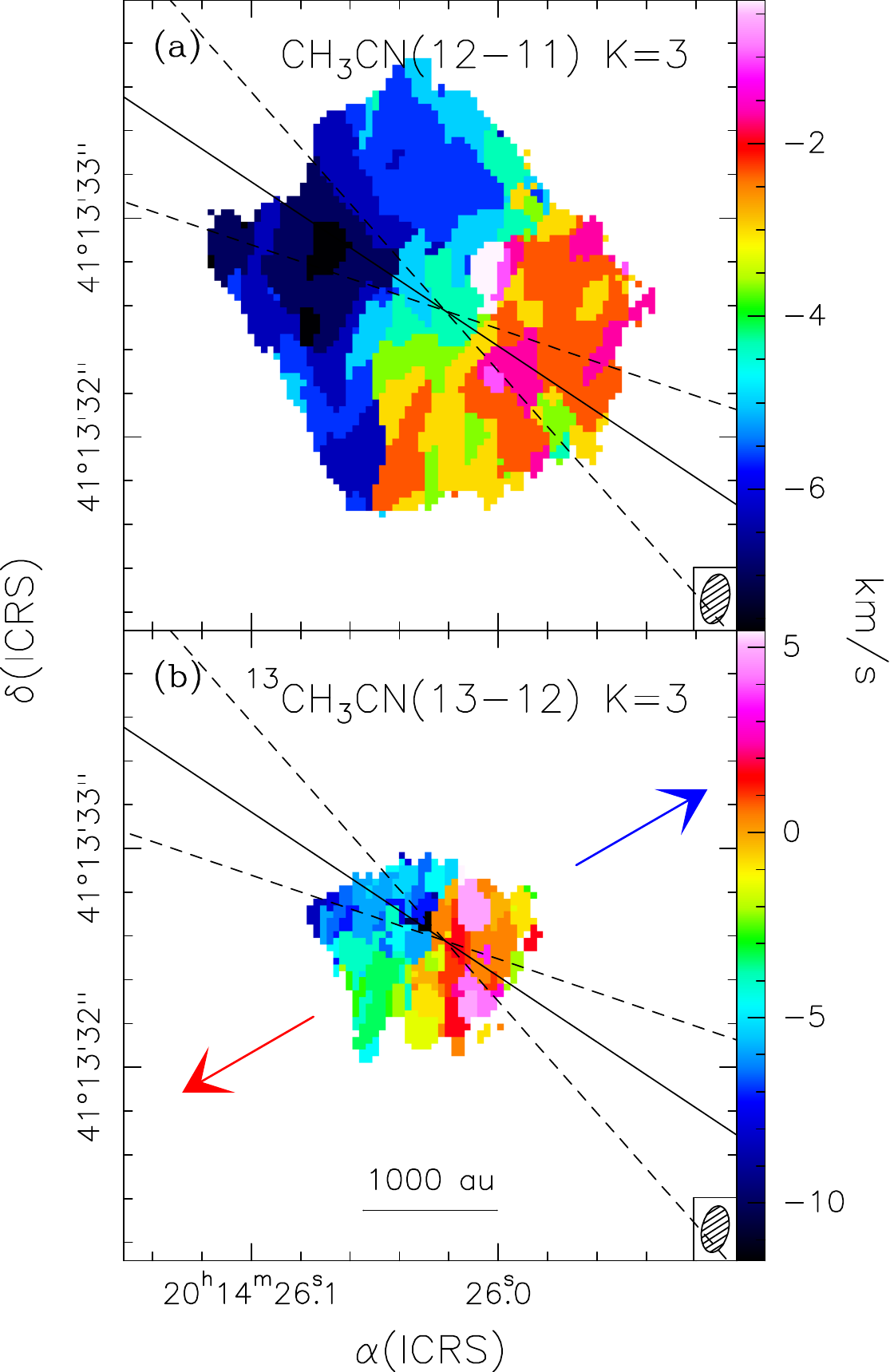}}
\caption{
Moment-9 maps of the \MCN(12--11) $K$=3 (panel {\bf a}) and \MCNII(13--12)
$K$=3 (panel {\bf b}) lines.
The systemic velocity is \mbox{$\sim$--3.9~\kms.}
The solid line shows the deconvolved direction
of the major axis of the continuum emission, while the dashed lines give the
errors on this direction. The blue and red arrows indicate the directions
of the blue- and redshifted lobes of the jet shown in Fig.~\ref{fjmaps}.
The ellipses in the bottom right corners denote the synthesized beams.
}
\label{fmomn}
\end{figure}

In order to explore the velocity field of the gas, we also show the moment-9
maps in Fig.~\ref{fmomn}, where for each pixel one plots the velocity of
the spectral channel where the maximum emission (i.e., the moment~8) is
attained. The new data resolve the structure of the disk and show that the
NE--SW velocity gradient first revealed by Cesaroni et al.~(\cite{cesa97})
is oriented parallel to the major axis of the continuum emission and is
preserved down to scales of as small as $\sim$200~au. We also note that the
range of peak velocities
spanned by the optically thick isotopolog ($\sim$7~\kms) is
narrower
than that of the optically thinner $^{13}$C
substituted species ($\sim$17~\kms). This finding is consistent with the idea that the
latter is tracing a more internal region of a Keplerian disk, where the
rotation velocity increases for decreasing radii. The kinematics of the
disk is further investigated in Sect.~\ref{sadisk}.

\section{Data analysis}
\label{sdis}

Although the disk and the associated jet (or outflow) in a YSO are tightly
connected entities, for the sake of clarity, we analyze them separately,
with our main aim being to shed light on their physical parameters and kinematical
structure.

\subsection{The jet}
\label{sajet}

The higher sensitivity and angular resolution with respect to previous
SiO observations of the jet (Su et al.~\cite{su07}) allow us to improve our
knowledge of its structure and velocity field. For this purpose, we adopt the
model by Cesaroni et al.~(\cite{cesa99}), where the bipolar jet is described
as a cone where the material is expanding radially with velocity proportional
to the distance from the star. After fixing the PA of the jet axis to
$\PAj$=--60\degr, the free parameters of the model are the opening angle
of the cone, $\theta$, the angle between the axis of the cone and the
plane of the sky, $\phi$, and the velocity gradient $V_0/R_0$. A more
detailed description is given in Appendix~\ref{appa}. As done by Cesaroni
et al.~(\cite{cesa99}), we only use the model to derive the borders of
the region in space and velocity where SiO emission is present. We change
the three input parameters of the model until these borders encompass the
observed emission, thus deriving the main parameters of the flow.

\begin{figure}
\centering
\resizebox{8.5cm}{!}{\includegraphics[angle=0]{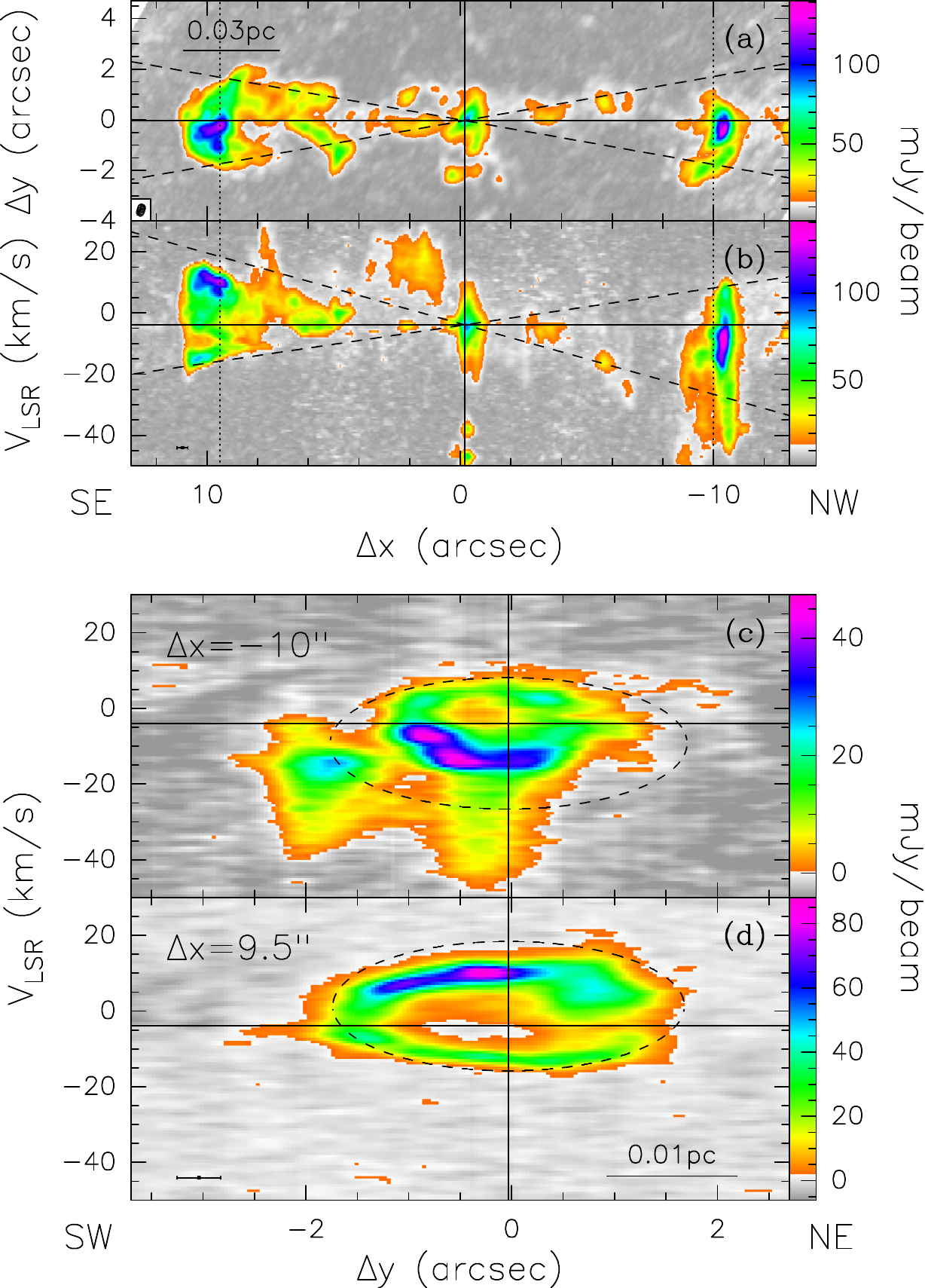}}
\caption{
 SiO(5--4) line emission in \I. The offsets are relative to
the phase center of the observations.
{\bf a.} Moment-8 map of the emission rotated by 30\degr\ clockwise,
so that the jet axis lies parallel to the X-axis of the plot. The dotted
lines indicate the cuts used to make the PV diagrams in panels (c) and (d).
According to our conical jet model, emission can be seen only in the region
comprised between the two dashed lines. The horizontal and vertical lines
mark the position of the 1.4~mm continuum peak.
{\bf (b)} Moment-8 PV diagram along the jet axis. The dotted lines have the
same meaning as in panel (a). The cross in the bottom left corner indicates
the angular and spectral resolutions.
The vertical line marks the position of the continuum peak
and the horizontal line the systemic velocity of --3.9~\kms.
{\bf (c)} PV diagram along the dotted line at offset +9\farcs5 in panel (a).
The dashed ellipse marks the region inside which emission is expected
according to our conical jet model.
The horizontal and vertical lines have the same meaning as in panel (b).
{\bf (d)} PV diagram along the dotted line at offset --10\arcsec\ in panel (a).
The dashed ellipse has the same meaning as in panel (c). The cross in the
bottom left corner denotes the angular and spectral resolutions.
The horizontal and vertical lines have the same meaning as in panel (b).
}
\label{fpvsio}
\end{figure}

While Cesaroni et al.~(\cite{cesa99}) constrained the model parameters
using only the map of the jet and the PV diagram along the jet axis, we
also use
the PV diagrams perpendicular to the jet axis across the two bow shocks
to the SE and NW. The result is shown in Fig.~\ref{fpvsio}, where the map
and PV plots of the SiO(5--4) line are compared to the borders obtained
from the best-fit model. In practice, the best fit was obtained using the
values derived by Cesaroni et al.~(\cite{cesa99}) as initial guesses
and then varying these until a good match with the map and PV diagrams
was found by eye. The resulting values are $\phi$=3\degr, $\theta$=10\degr,
and $V_0/R_0$=10~\kms\,arcsec$^{-1}$=1.3~\kms\,pc$^{-1}$. The
only significant difference with respect to the values obtained by
Cesaroni et al.~(\cite{cesa99}) from lower resolution observations
of the SiO(2--1) line ($\phi$=9\degr, $\theta$=21\degr, and
$V_0/R_0$=8.3~\kms\,arcsec$^{-1}$=1.3~\kms\,pc$^{-1}$) is that the jet
appears to be more collimated by a factor of $\sim$2.

The PV plots across the bow shocks at the two ends of the jet (see bottom
panels of Fig.~\ref{fpvsio}) reveal that the SiO emission is indeed confined
to the outer layer, and thus seems to be tracing the part of the cavity walls
opened by the jet that corresponds to the post-shock region.

\begin{figure}
\centering
\resizebox{8.5cm}{!}{\includegraphics[angle=0]{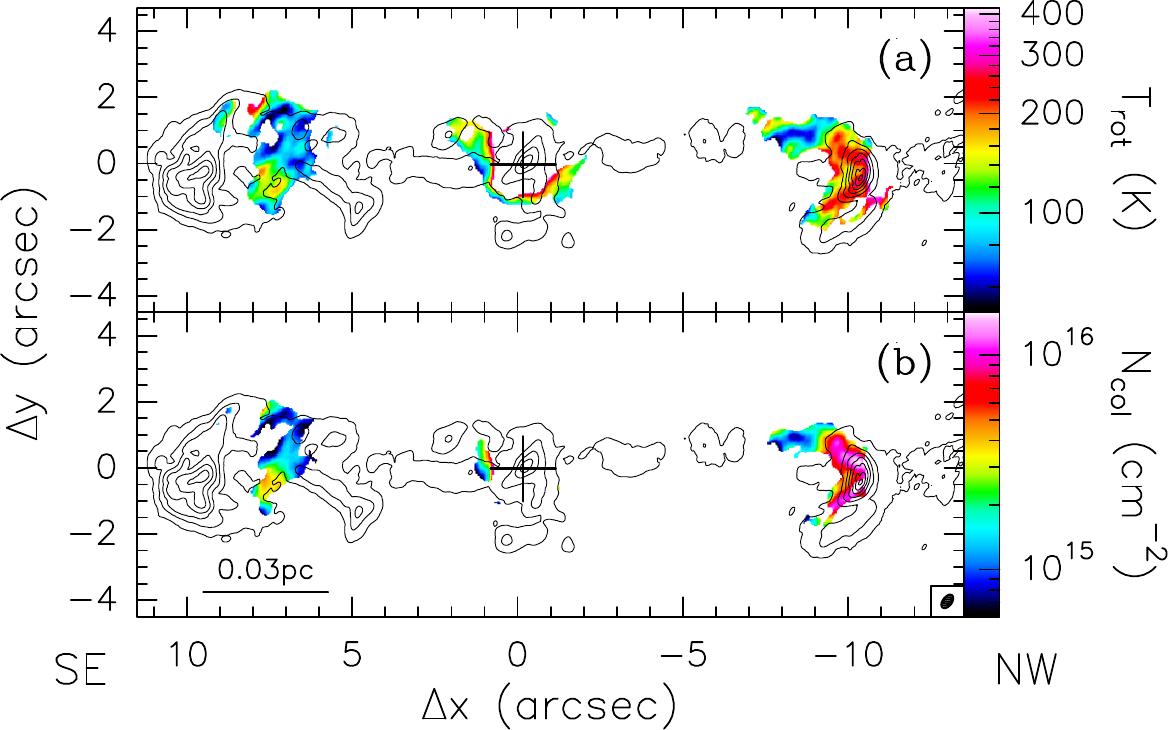}}
\caption{
Contour map of the SiO emission (same as in Fig.~\ref{fjmaps}a) overlaid
on maps of the \FORM\ rotational temperature (panel {\bf a})
and column density (panel {\bf b}). The maps have been rotated by 30\degr\
clockwise. The ellipse in the bottom right corner denotes the synthesized
beam. The cross marks the position of the 1.4~mm continuum peak.
}
\label{ftn}
\end{figure}

Further analysis of the jet can be performed by means of the \FORM\ lines.
Given the different energies of the three lines (see Sect.~\ref{srjet}),
we could obtain maps of the rotational temperature, $T_{\rm rot}$, and
column density, $N_{\rm col}$, of the \FORM\ molecule, assuming optically
thin emission and local thermodynamic equilibrium (LTE). For this purpose,
we used the maps of the integrated emission over the lines. In particular,
we computed the ratio between the map of the ($3_{0,3}$--$2_{0,2}$) line
and that obtained by averaging the maps of the ($3_{2,2}$--$2_{2,1}$) and
($3_{2,1}$--$2_{2,0}$) lines. The results are shown in Fig.~\ref{ftn},
where only values with relative errors of $<$60\% are displayed. We note
that the method does not provide us with reliable values close to the
(proto)star, where the emission is likely optically thick, and in part of
the bow shocks, where the emission is too faint or the temperature is too
high to be measured with transitions whose energies lie below 68~K. Despite
these limitations, one sees that the bow shock to the NW presents a clear
structure, where the maximum temperature and column density are attained
along the leading edge, while the colder gas lies in the tail. As for the
bow shock to the SE, \FORM\ emission is detected only in the
trailing
wake, which appears cooler and less dense. As a whole, this picture reveals differences in the evolution of the two lobes of the jet, where the NW lobe is
still impinging on dense material, while the SE lobe has already
pierced its way through a lower density medium.

\subsection{The disk}
\label{sadisk}

Previous studies of \I\ (e.g., Palau et al.~\cite{palau15}) have shown that the
disk is detected in COMs, the typical tracers
of hot molecular cores (HMCs); although some emission in the same molecules
is also detected from shocks at the base of the outflow.  Our frequency
setup covers many lines of COMs, but in the present article we focus only
on methyl cyanide and its isotopologs (\MCNI\ and \MCNII) and on the
DCN(3--2) line, which appears to trace the disk over larger radii than \MCN.

\subsubsection{Evidence for Keplerian rotation}
\label{skep}

\begin{figure}
\centering
\resizebox{8.5cm}{!}{\includegraphics[angle=0]{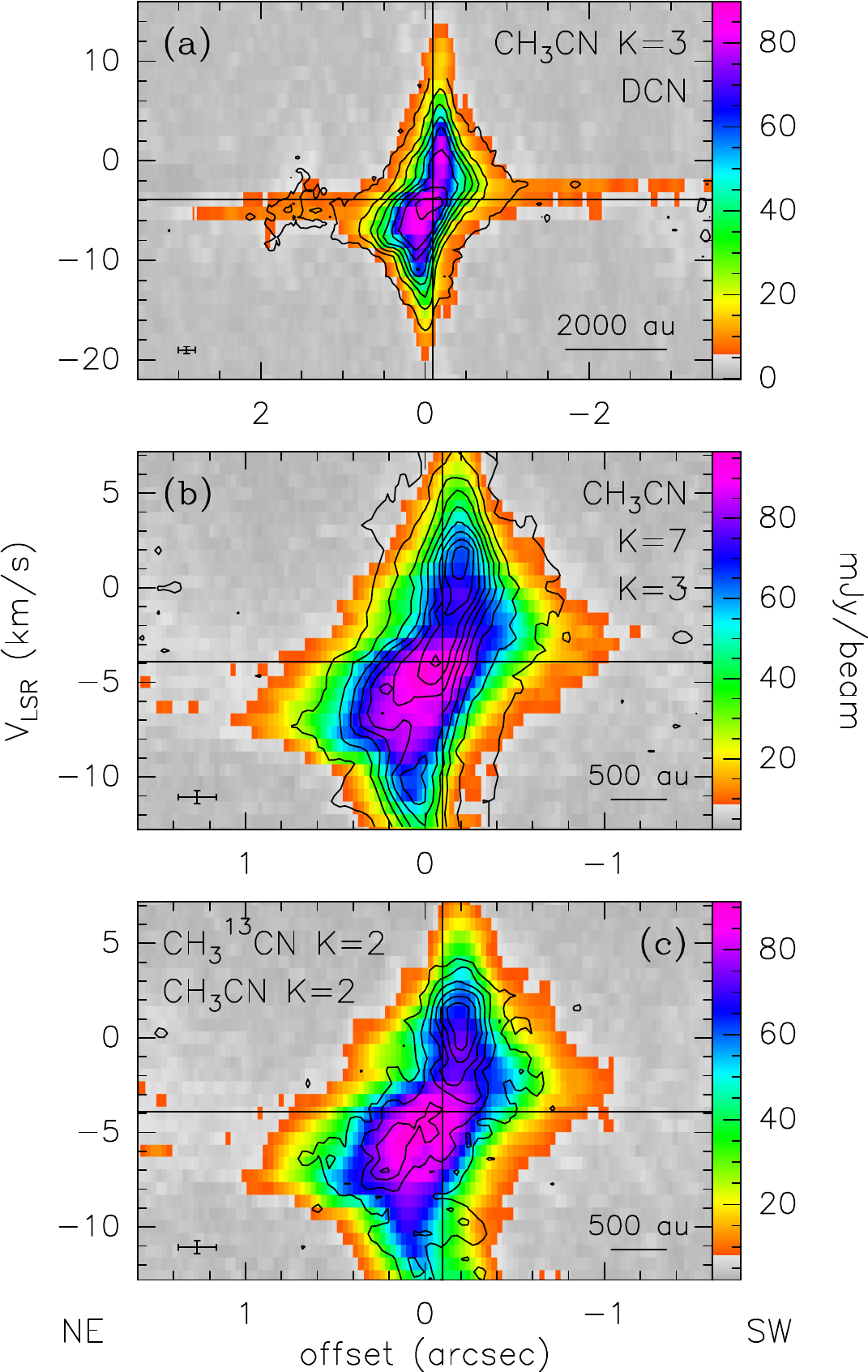}}
\caption{
Moment-8 PV diagrams of various molecular transitions along the disk plane
(PA=53\degr). The offset is relative to the phase center of the observations.
The horizontal and vertical lines mark the
systemic velocity and the position of the 1.4~mm continuum peak, respectively.
The crosses in the bottom left corners indicate the angular and spectral
resolutions.
{\bf (a)} PV plot of the \MCN(12--11) $K$=3 component (contours) overlaid
on that of the DCN(3--2) line. Contour levels range from 7 to 91 in steps
of 12~mJy/beam.
{\bf (b)} Same as panel (a), with \MCN(12--11) $K$=7 overlaid on $K$=3.
Contour levels range from 5 to 53 in steps of 6~mJy/beam.
{\bf (c)} Same as panel (a), but with \MCNI(12--11) $K$=2 overlaid on \MCN(12--11)
$K$=2.  Contour levels range from 5 to 20 in steps of 3~mJy/beam.
}
\label{fdpvs}
\end{figure}

\begin{figure}
\centering
\resizebox{8.5cm}{!}{\includegraphics[angle=0]{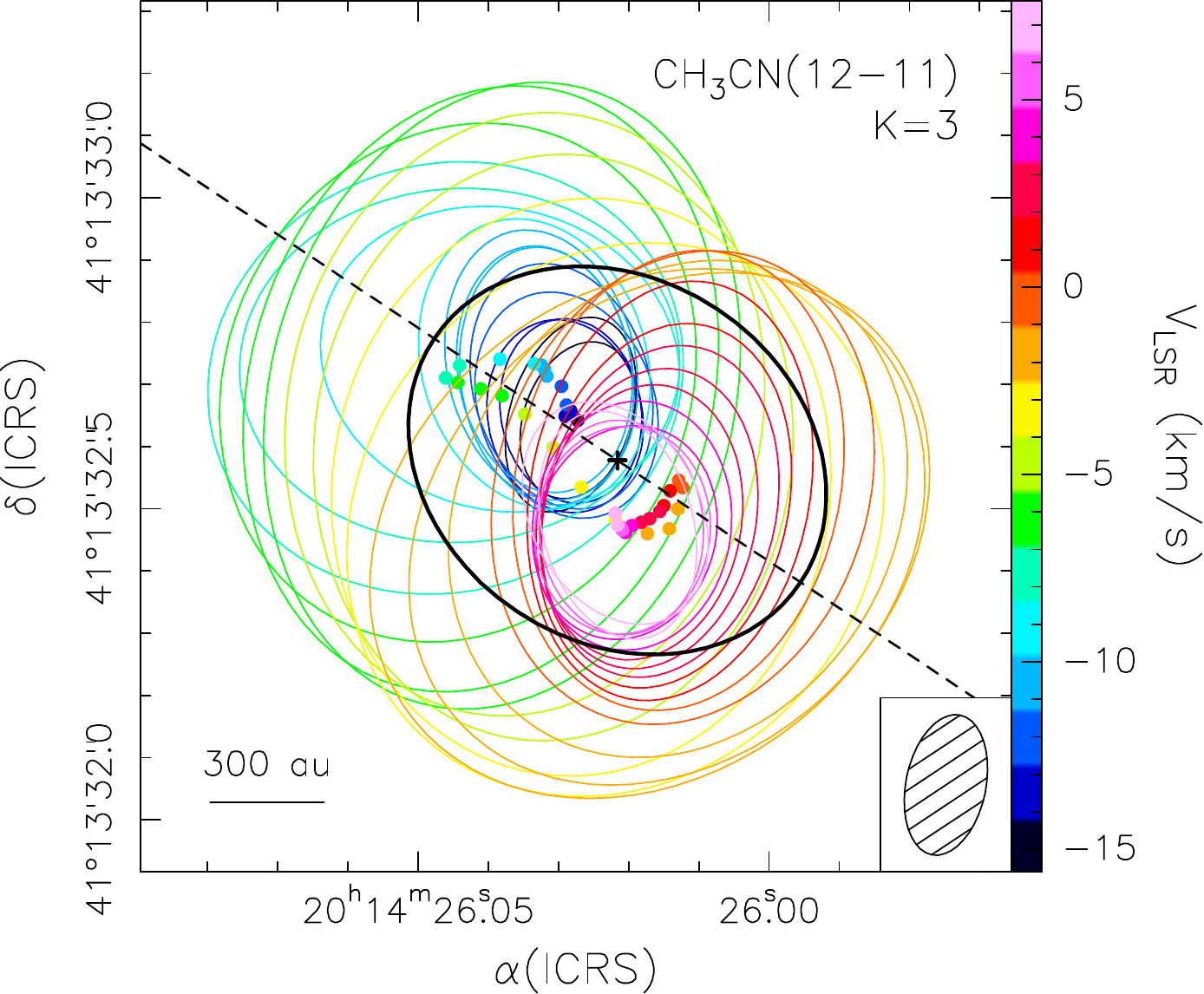}}
\caption{
Distribution of the peaks (colored points) obtained by fitting a 2D Guassian
to the emission in each channel of the channel maps of the \MCN(12--11)
$K$=3 line. The colored ellipses are the deconvolved FWHM of the fitted
Gaussians. The color corresponds to the LSR velocity as indicated by the
scale to the right. The black cross and ellipse represent
the peak position and deconvolved FWHM of the 1.4~mm continuum emission, respectively. The
dashed line outlines the direction of the major axis of the black ellipse.
The shaded ellipse in the bottom right corner represents the synthesized beam.
}
\label{ffwhp}
\end{figure}

In Fig.~\ref{fdpvs}, we compare the PV diagrams along the disk plane
in different lines. The PA of the plane of the disk was estimated to be
$\PA$$\simeq$$53\degr$ (Cesaroni et al.~\cite{cesa05}) and we assume this value
for our analysis. As noted by Cesaroni et al.~(\cite{cesa05,cesa14}),
the pattern of the PV plots has a ``butterfly'' shape, strongly suggestive
of Keplerian rotation. To further support this interpretation, we
fitted the emission in each channel of the \MCN(12-11) $K$=3 component
with a 2D Gaussian and in Fig.~\ref{ffwhp} we plot the peak positions thus
obtained. We also draw ellipses corresponding to the deconvolved FWHM of
the Gaussian fit. It is interesting to note that the orientation of the
ellipses changes systematically going from the high velocities to the low
velocities, with the latter ellipses oriented SE--NW and the former NE--SW.

\begin{figure}
\centering
\resizebox{8.5cm}{!}{\includegraphics[angle=0]{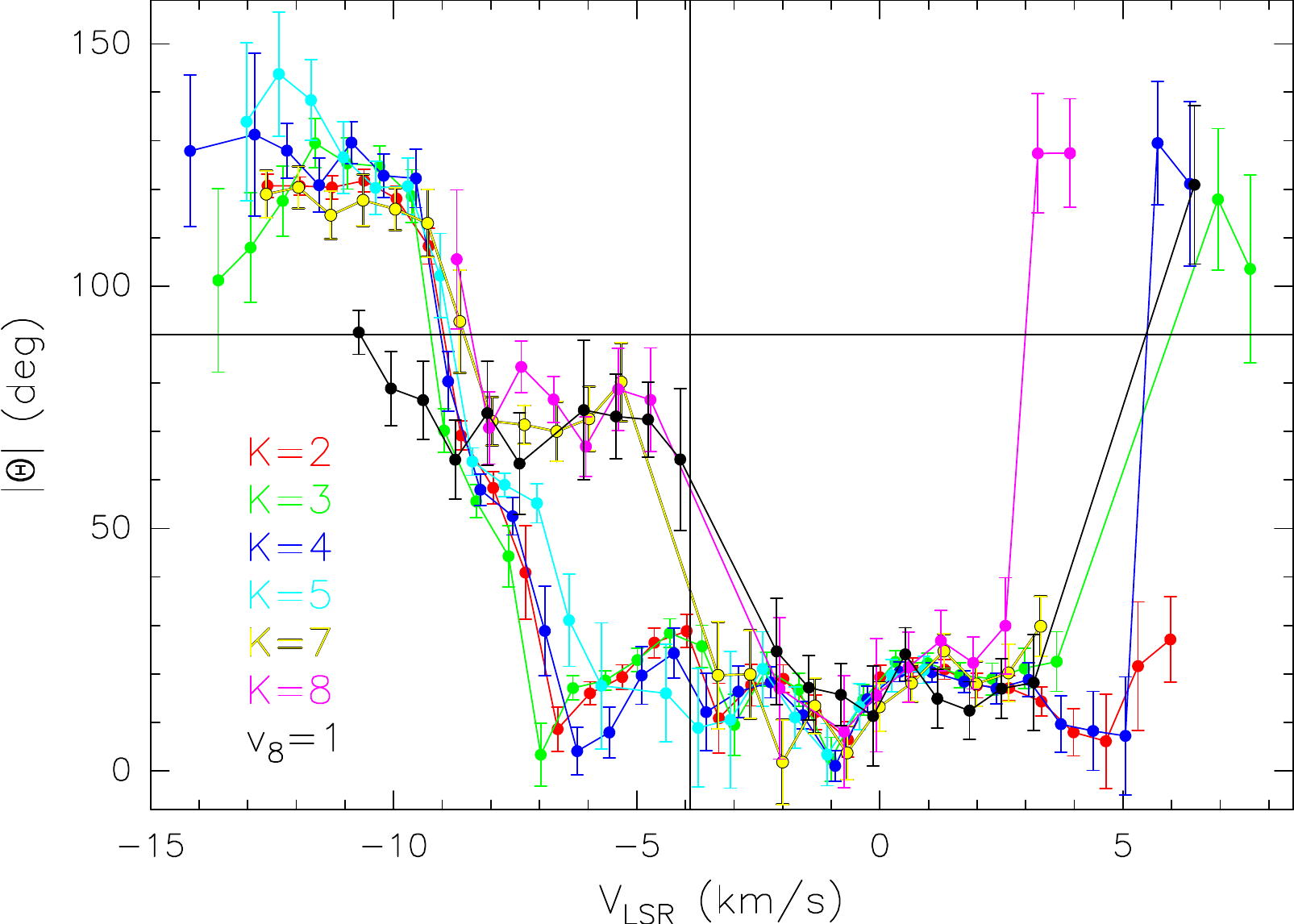}}
\caption{
Absolute value of the angle between the major axis of the deconvolved FWHM
of the emission in the line channel, and the axis of the disk versus the
LSR velocity of the channel. The colors correspond to the \MCN(12--11)
lines as indicated in the figure. The vertical line marks
the systemic velocity. The horizontal line corresponds to $|\Theta|=90\degr$.
}
\label{fpav}
\end{figure}

\begin{figure}
\centering
\resizebox{8.5cm}{!}{\includegraphics[angle=0]{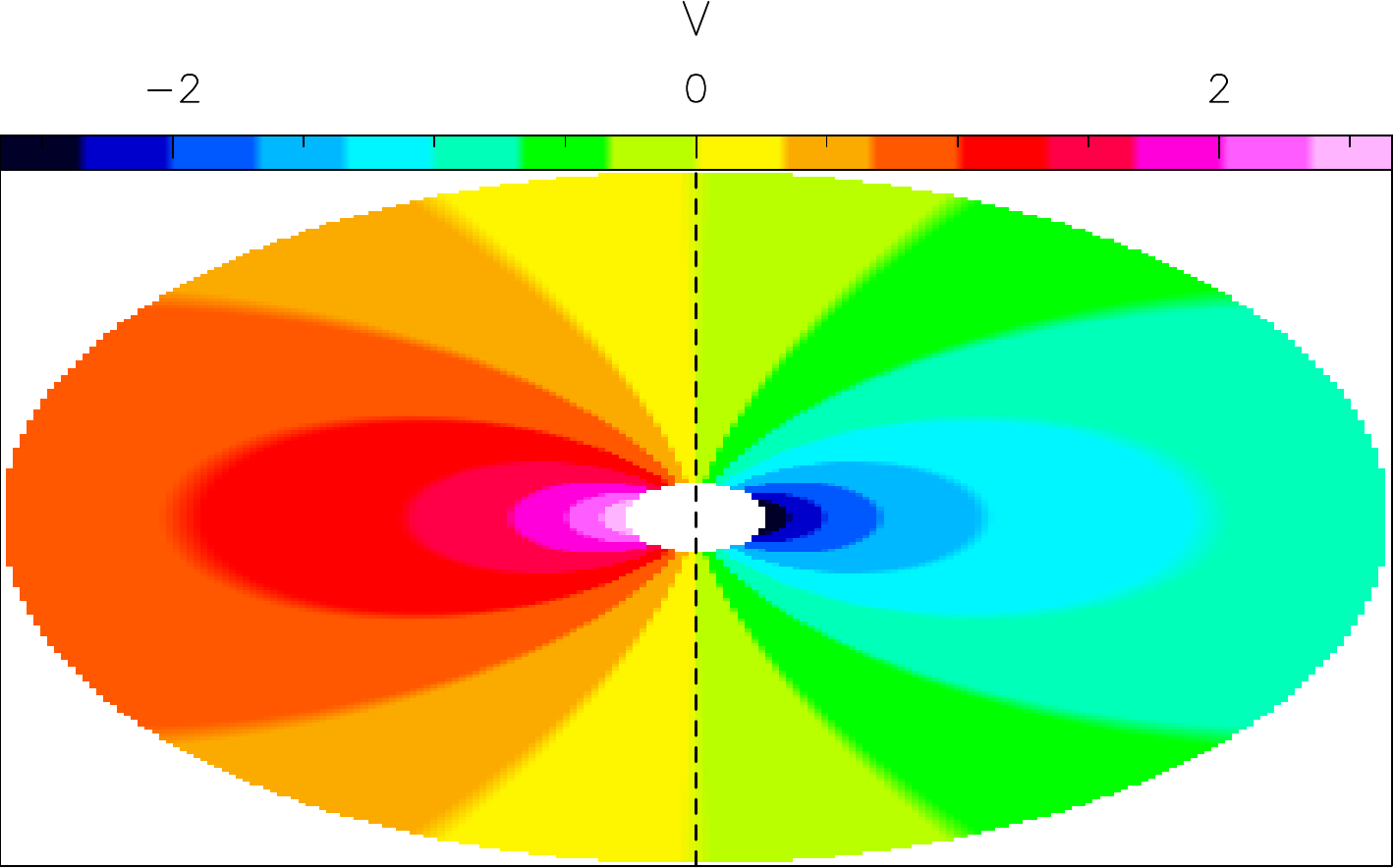}}
\caption{
Schematic plot of the velocity along the line of sight relative to the
star velocity, in a template Keplerian disk inclined by 30\degr. The
plane of the figure is the plane of the sky and the dashed line is the
projection of the disk rotation axis. The color scale indicating the
velocity is in arbitrary units. We note that the high-velocity magenta/red
and blue patterns are elongated perpendicular to the yellow/green patterns,
which correspond to velocities close to zero.
}
\label{fvpat}
\end{figure}

This effect is better visualized in Fig.~\ref{fpav}, where we plot
the absolute value of the angle $\Theta$ between the major axis of the
ellipses and the disk rotation axis, which is assumed to be perpendicular to the major
axis of the continuum emission. This is done for the strongest and least
blended lines indicated in the figure. By definition, $\Theta$=90\degr\ for
the continuum emission. The plot shows that, at high velocities, $|\Theta|$
is close to 90\degr, whereas close to the systemic velocity, $|\Theta|$ tends
to 0\degr. Such behavior is expected in a Keplerian disk, as one can
see from the template velocity pattern in Fig.~\ref{fvpat}, where the
high-velocity red- and blueshifted emission arises from elongated regions
perpendicular to the disk rotation axis, whereas the systemic velocity
traces regions parallel to it.

\subsubsection{Derivation of \Trot\ and \Ncol}
\label{stn}

To estimate some physical parameters of the disk, we rely on a large
number of methyl cyanide lines with different excitation energies, which can be
analyzed under the LTE approximation given the large densities ($>10^7$~\cmc;
see e.g., Kurtz et al.~\cite{ppiv}) typical of HMCs.

\begin{figure}
\centering
\resizebox{8.5cm}{!}{\includegraphics[angle=0]{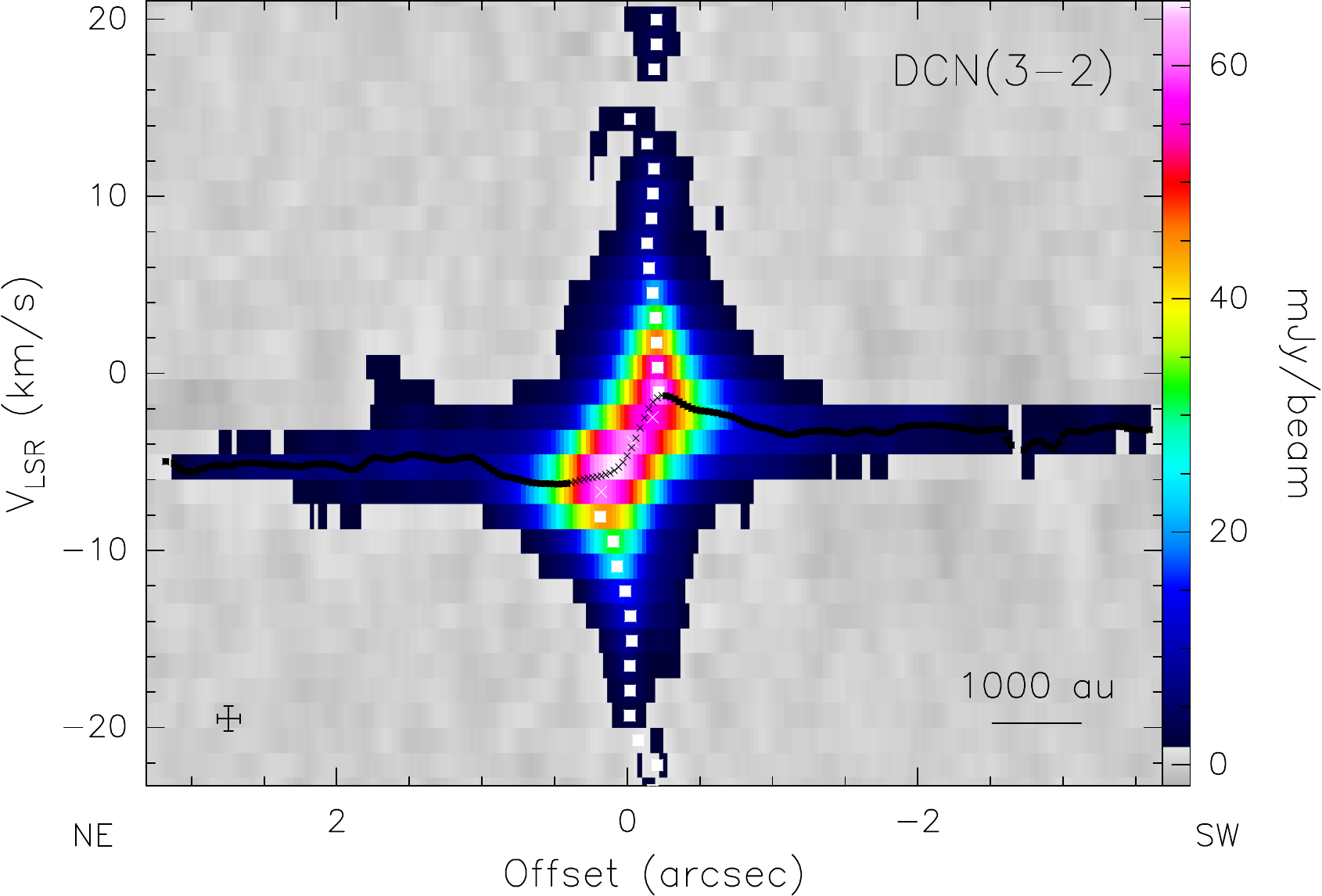}}
\caption{
Moment-8 PV diagram of the DCN(3--2) line emission along the plane of the
disk (PA=53\degr). The offset is relative to the phase center of the
observations. The black and white points mark the emission peaks
obtained by fitting a Gaussian to the emission, respectively, for fixed
offset and fixed velocity. The solid squares indicate the points selected
by us to define the velocity--radius relation in the Keplerian disk.
}
\label{fpvdcn}
\end{figure}

Thanks to the high resolution and sensitivity of our observations, we
can improve on previous estimates of the temperature and column density
(Cesaroni et al.~\cite{cesa97}), which were only mean values over the whole
disk. In particular, we wish to obtain the \MCN\ rotational temperature,
\Trot, and column density, \Ncol, as a function of radius in the disk. To this end, we use the PV diagrams of selected methyl cyanide lines with
the approach described below.

Since the disk is undergoing Keplerian rotation, the emission in the PV
plots is expected to have a ``butterfly'' shape, where the outer border
of the pattern (i.e., the hyperbolic branches with the highest and lowest
velocities) corresponds to the velocities at the intersection between
the disk and the plane of the sky. Therefore, the emission along these
two branches can be used to compute the physical parameters as a function
of radius. The latter is simply the offset in the PV plot relative to the
continuum position. However, the observed data result from the convolution
of the butterfly-shaped pattern with the synthesized beam and the
intrinsic line width, which makes it non-trivial to identify the above-mentioned hyperbolic branches. To overcome this problem, we decided to use an
empirical method. We fitted the emission in the PV plot of the DCN line with
a Gaussian, both for fixed velocity and for fixed offset, thus deriving,
respectively, the offset and the velocity corresponding to the peak
emission. We note that we chose the DCN line because this is detected
at larger disk radii than the other transitions observed by us.
The distribution of these peaks is shown in Fig.~\ref{fpvdcn},
where the white and black points indicate the peaks obtained
with fixed velocity and fixed offset, respectively. With the exception of the central
region of the plot
(crosses),
the other points (solid squares)
describe two hyperbolic branches that we can use for our analysis.

\begin{figure*}
\resizebox{8.3cm}{!}{\includegraphics[angle=0]{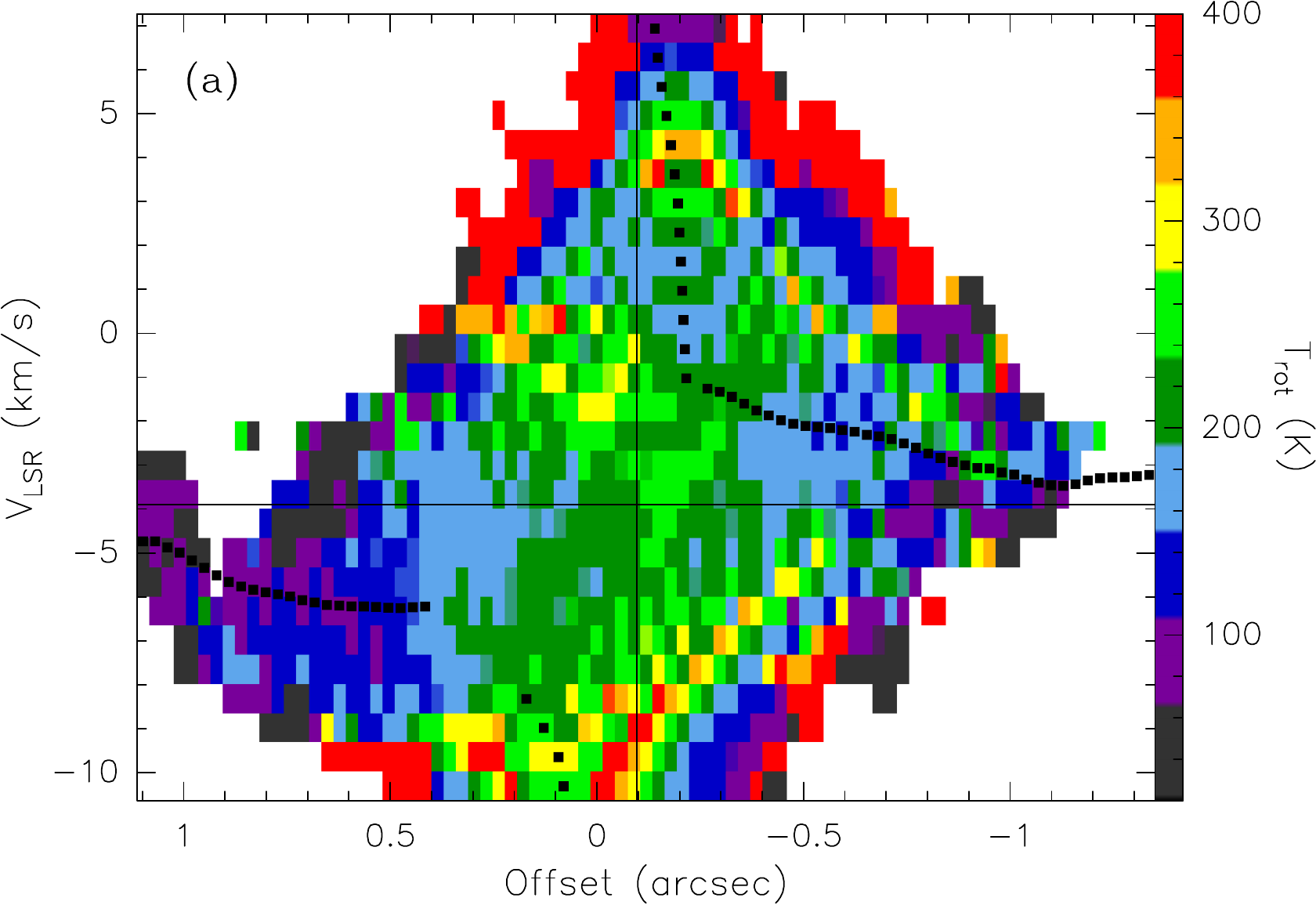}} \hspace*{2mm}
\resizebox{8.3cm}{!}{\includegraphics[angle=0]{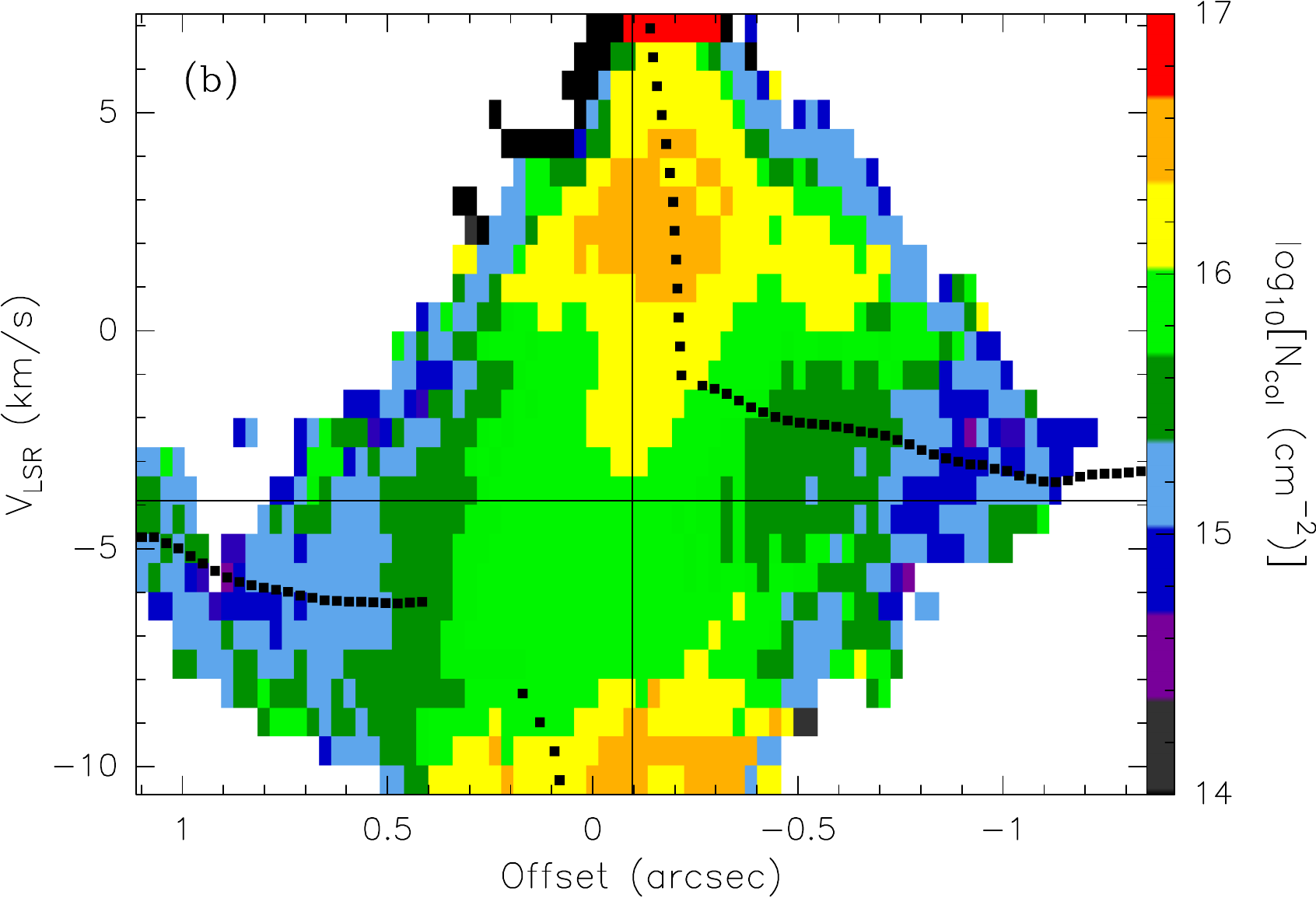}} \\[2mm]
\resizebox{7.8cm}{!}{\includegraphics[angle=0]{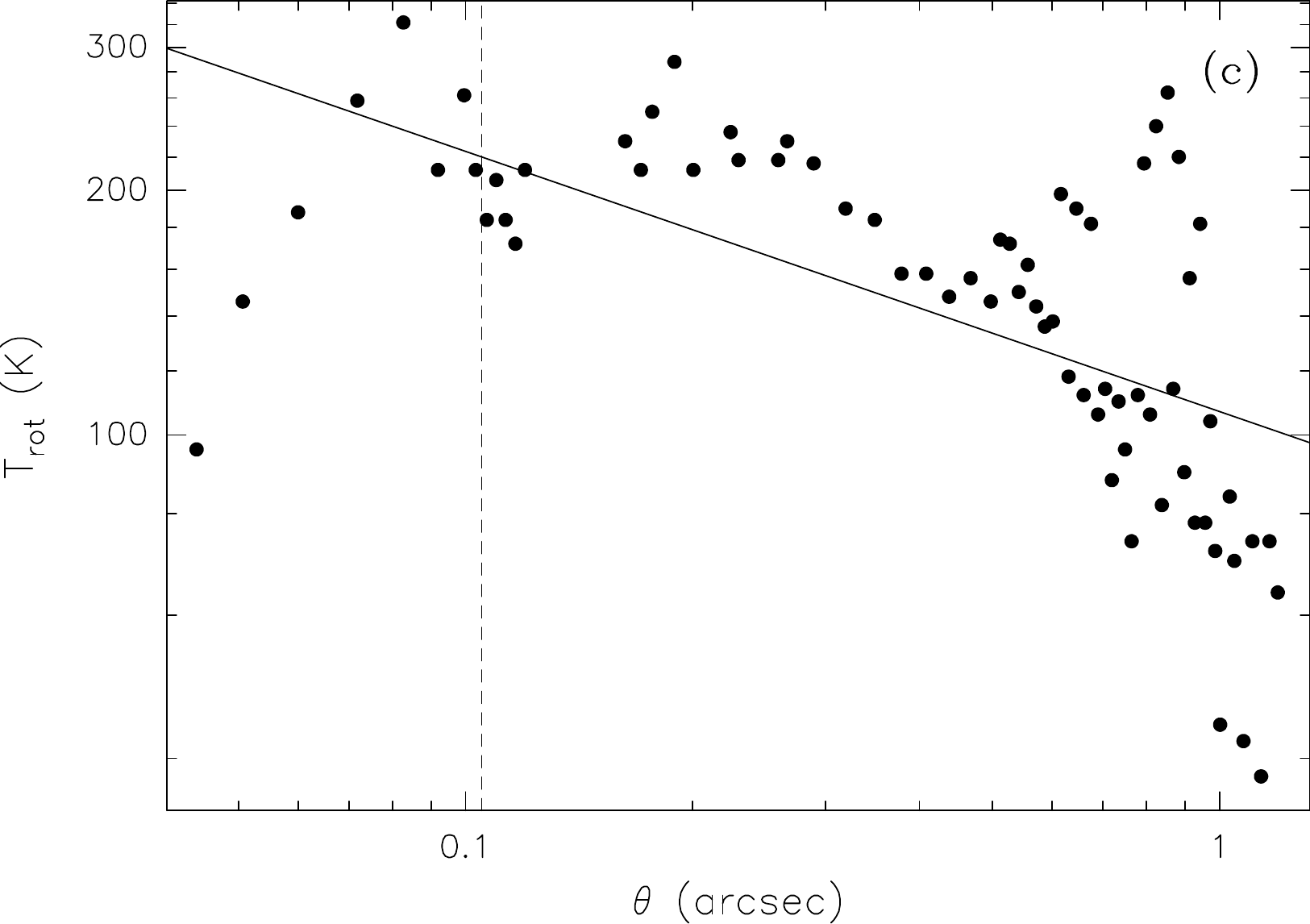}} \hspace*{4mm}
\resizebox{7.8cm}{!}{\includegraphics[angle=0]{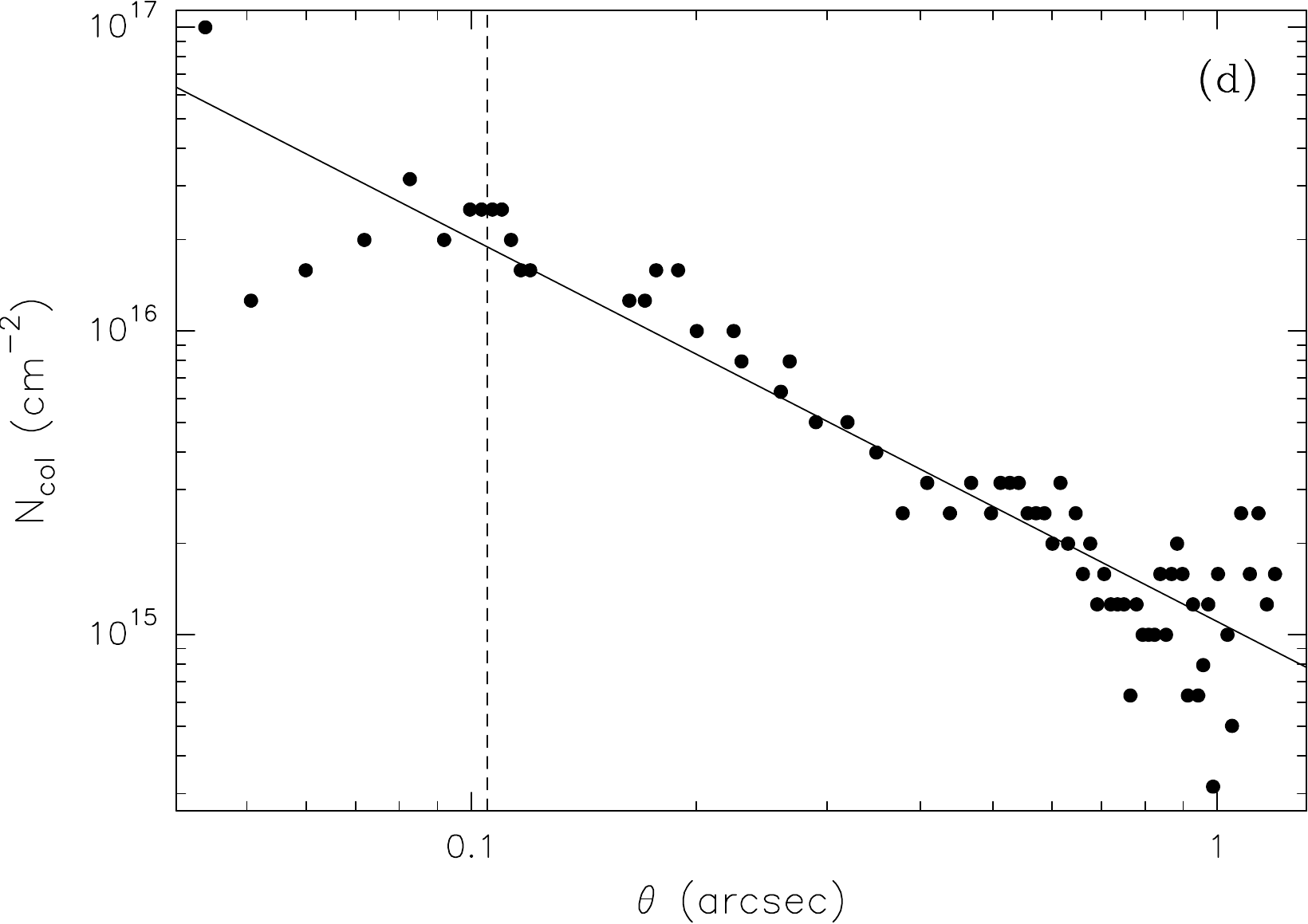}}
\caption{
Analysis of the temperature and column density distribution in the disk
around \I.
({\bf a)} PV diagram along the disk plane of the \MCN\ rotation temperature
obtained by fitting the brightness temperature of selected transitions.
The black solid squares mark the points used to derive \Trot\ as a function
of radius (see text).
{\bf (b)} Same as panel (a) but for the \MCN\ column density.
{\bf (c)} Rotation temperature versus angular radius obtained from the points of the PV
diagram marked by the black solid squares. The solid line is a linear
fit to the data corresponding to the relation
$\Trot({\rm K})=107~[\theta({\rm arcsec})]^{-0.32}$. The dashed line corresponds
to half the synthesized beam.
{\bf (d)} Same as panel (c) but for the \MCN\ column density. The
relationship obtained from the linear fit to the data is
$\Ncol({\rm cm^{-2}})=1.1\times10^{15}~[\theta({\rm arcsec})]^{-1.3}$.
}
\label{fpvtn}
\end{figure*}

For our calculation of \Trot\ and \Ncol, we selected only the lines of the
main species that were least affected by blending with other transitions,
namely the $K$=2,3,4,5,7,8 components and the $v_8$=1 $K,l$=2,1 line.
In cases where the line wings were overlapping with a nearby line, the channels
affected by this blending were clipped. For each of these lines, a PV plot
with the same grid of pixels was built and the line brightness temperature,
\Tb, in each of the solid points in Fig.~\ref{fpvdcn} was extracted. All
these values of \Tb\ were fitted with the expression
\begin{equation}
 T_{\rm B}^{\rm mod} = \eta \, T_{\rm rot} \left(1-\e^{-\tau}\right)
,\end{equation}
where $\eta\le1$ is a filling factor and the line opacity is given by
\begin{equation}
 \tau = \frac{1}{4\pi\delta V} \sqrt{\frac{\ln2}{\pi}} \frac{A_{ul}c^3}{\nu^3}
        g_u \frac{N_{\rm col}}{Q(T_{\rm rot})} \e^{-\frac{E_u}{kT_{\rm rot}}}
        \left(\e^{\frac{h\nu}{kT_{\rm rot}}}-1\right)
,\end{equation}
where $A_{ul}$ is the Einstein coefficient, $\delta V$ the channel width in velocity,
$c$ the speed of light, $\nu$ the line frequency, $Q(\Trot)$ the partition function, $E_u$
the upper level energy, $h$ the Planck constant, and $k$ the Boltzmann constant. Here, we assume LTE and the best fit was obtained by minimizing the expression
$\sum_i\left(T_{{\rm B}i}-T_{{\rm B}i}^{\rm mod}\right)^2$ , where the sum
runs over the selected \MCN\ lines (see above). For this purpose, the free
variables $\eta$, \Trot, and \Ncol\ were varied over reasonable intervals. In
this way, we obtained \Trot\ and \Ncol\ of \MCN\ at each pixel of the PV plot,
as shown in Figs.~\ref{fpvtn}a and~\ref{fpvtn}b. The calculation was done
only for the pixels of the PV diagram where at least three lines were detected.

Figures~\ref{fpvtn}c and~\ref{fpvtn}d show \Trot\ and \Ncol\ as functions
of the angular radius $\theta$
obtained directly from the offset in Figs.~\ref{fpvtn}a and~\ref{fpvtn}b.
Finally, from linear fits to the data, we derive the relations
\begin{equation}
 T_{\rm rot}({\rm K}) = 107~\left[\theta({\rm arcsec})\right]^{-0.32} = 1140~\left[R({\rm au})\right]^{-0.32}   \label{etrot}
,\end{equation}
\begin{equation}
 N_{\rm col}({\rm cm^{-2}}) = 1.1\times10^{15}~\left[\theta({\rm arcsec})\right]^{-1.3} = 1.7\times10^{19}~\left[R({\rm au})\right]^{-1.3}.   \label{encol}
\end{equation}

\begin{figure}
\centering
\resizebox{8.5cm}{!}{\includegraphics[angle=0]{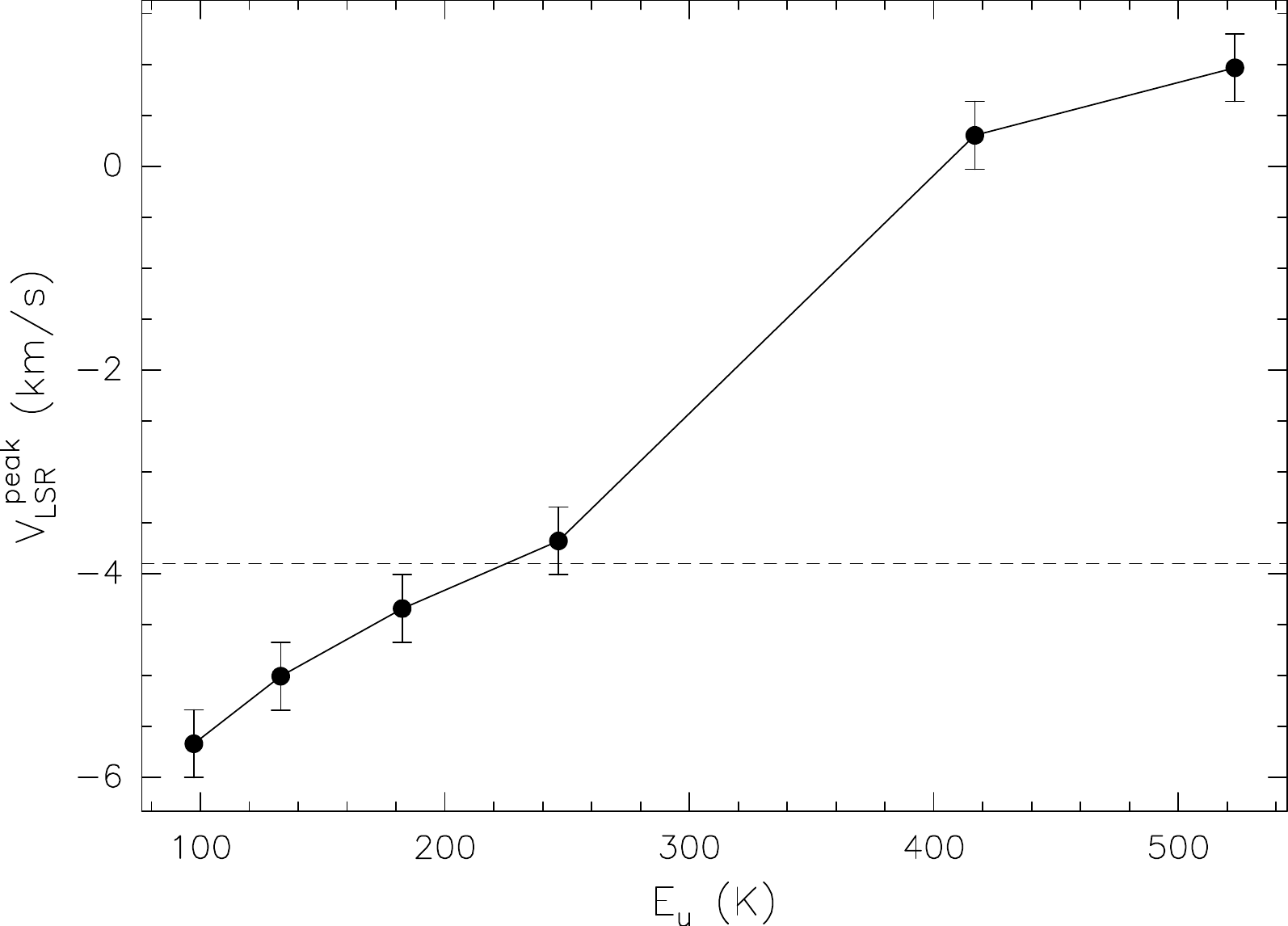}}
\caption{
 LSR velocity corresponding to the peak emission in the PV
diagrams of the \MCN(12--11) $K$=2,3,4,5,7,8 lines as a function of the
upper energy of the transition.
The dashed line marks the systemic LSR velocity of --3.9~\kms.
}
\label{fpvmax}
\end{figure}

\subsubsection{Evidence for infall}

In addition to rotation, the presence of accretion in the disk is suggested by
the shape of the PV plots, because
for Keplerian rotation no emission should be seen either at redshifted
velocities and positive offsets (with respect to the position of the star)
or at blueshifted velocities and negative offsets. This is clearly not
true for the PV diagrams in Fig.~\ref{fdpvs}. It must be noted that the
emission in these two quadrants of the PV plots cannot be explained with
smearing due to the finite spectral and angular resolution. If this were
the case, in Fig.~\ref{fdpvs}c, the emission at offset --0\farcs33 and
velocity --7~\kms\ , for example, would be attenuated by $3\times10^{-3}$ times in space
(for HPBW$\simeq$0\farcs16) and $5\times10^{-2}$ times in velocity (assuming
an intrinsic line FWHM=3~\kms). These factors are too small to allow for
the intensity of 22~mJy/beam measured at that point, taking into account
the fact that the maximum intensity in the PV plot is 97~mJy/beam.
The effect of adding a radial velocity component to Keplerian rotation is
described by Eq.~(1) of Cesaroni et al.~(\cite{cesa11}); in particular,
their Fig.~13 illustrates the pattern expected in the PV diagram.

Further possible evidence of infall comes from a noticeable feature of the PV
plots in Fig.~\ref{fdpvs}b and~\ref{fdpvs}c, where we note a striking
difference between the peak of the emission in lines with different
opacities and excitation energies. For instance, the \MCN\ $K$=2 component
peak is blueshifted, whereas the $K$=2 line of the \MCNI\ isotopolog and
the \MCN\ $K$=8 line peak at redshifted velocities. This behavior is
evident in Fig.~\ref{fpvmax}, where we plot the velocity of the peak in
the PV diagram as a function of the excitation energy of the transition
for the least blended \MCN\ lines with the best S/N. Clearly, the higher
the excitation energy, the more redshifted the emission. While several
explanations are possible, at the end of Sect.~\ref{smodd} we propose that
this trend is due to the presence of accretion.

\begin{figure}
\centering
\resizebox{8.5cm}{!}{\includegraphics[angle=0]{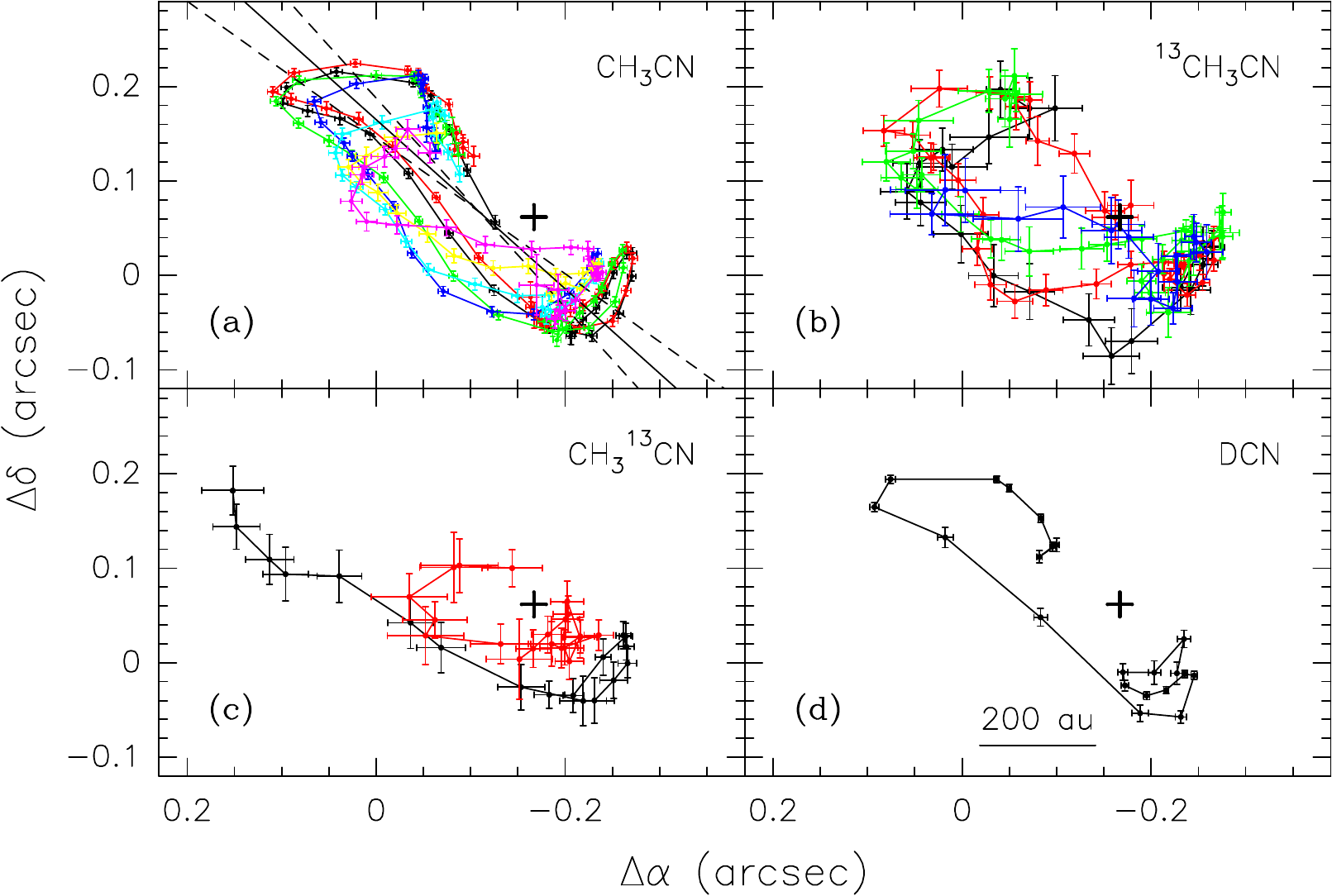}}
\caption{
Distributions of the peaks of the emission in the velocity channels
of various lines observed in \I. The error bars are the formal errors
on the Gaussian fits. The offsets are relative to the phase center of
the observations. The black cross marks the position of the 1.4~mm
continuum peak.
{\bf (a)} Plot obtained from the \MCN(12--11) lines. The color coding is as
follows: $K$=2 black, 3 red, 4 green, 5 blue, 7 cyan, 8 yellow, and
$v_8$=1 magenta. The dashed lines are linear fits to the points obtained by
fitting Y vs X and X vs Y; the solid line is the mean of the two.
{\bf (b)} Same as panel (a) but for the \MCNII(13--12) components: $K$=2 black, 3 red,
4 green, and 5 blue.
{\bf (c)} Same as panel (a) but for the \MCNI(12--11) components: $K$=2 black, 4 red.
{\bf (d)} Same as panel (a) but for the DCN(3--2) line.
}
\label{fvpeaks}
\end{figure}

\subsubsection{A model for the disk}
\label{smodd}

An intriguing feature of the disk in \I\ is the distribution of the
emission peaks in Fig.~\ref{fvpeaks}. As already noted by Cesaroni et
al.~(\cite{cesa14}), the positions of the peaks of the methyl cyanide
lines at different velocities are not aligned along a straight line
passing through the disk center, but describe a C-shaped structure. The
case of \I\ is not unique because the same C-shaped pattern has
also been found by S\'anchez-Monge et al.~(\cite{sanch13}) and
Bayandina et al.~(~\cite{baya24})
in two similar objects. Here, we propose a toy model that
can qualitatively explain the shape of these plots as well as the trend
of the peak velocity with energy of the line.

The model is intended to be a proof of concept to explain the results
illustrated in Figs.~\ref{ffwhp}, \ref{fvpeaks}, and~\ref{fpvmax}.
Briefly, we propose that such results are due to the presence of a dusty
envelope enshrouding the disk itself. We assume a simple scenario with
a geometrically thin disk
and a spherical halo extending from the center up to the outer radius of
the disk. The density of the halo is
$\propto r^{-1}$,
where $r$ is the distance from the disk center.
The disk is dust-free and optically thick in the line emission, whereas
the line emission and absorption in the dusty halo are negligible. Incidentally,
we note that, while the outflow carves a cavity in the spherical halo, it
is likely to be collimated enough not to affect the spherical assumption
for the radiative transfer modeling we perform.
We also assume that the disk is undergoing Keplerian rotation and
radial infall with velocity $\propto r^{-1/2}$. The model is described
in Appendix~\ref{appb}.

The best fit to the data is obtained in two steps: first, we fit the observed
LSR velocity of the disk as in S\'anchez-Monge et al.~(\cite{sanch13}),
thus determining all parameters except the envelope opacity; then we vary
this opacity until we reproduce the C-shaped distribution of the peaks. Below, we describe these two steps one by one.

\paragraph{Fit of the LSR velocity}
~\\[3mm]
To fit the disk velocity field, we minimize the expression
$\sum_i\left(V_i^{\rm obs}-V_i^{\rm mod}\right)^2$, where $V_i^{\rm obs}$ is
the LSR velocity of a peak in Fig.~\ref{fvpeaks} and $V_i^{\rm mod}$ is the LSR
velocity computed from Eq.~(\ref{eavlsr}) at the position of that peak. We
decided to use only the peaks of the \MCN\ ground-state $K$ components in
Fig.~\ref{fvpeaks}a, because these have the best S/N and consequently
the smallest positional errors. As explained in Appendix~\ref{avlsr}, the
input variables of the disk model are the coordinates of the star (\xs,\ys),
the systemic velocity (\Vs), the PA of the projected major axis
of the disk ($\PA$), the angle between the disk plane and the plane of the
sky ($\psi$), the mass of the star (\Ms), and the ratio ($A$) between the
radial (i.e., infall) velocity and the rotation velocity. We fixed $\PA$
to the value obtained from a least-square fit to the peaks of all ground-state $K$ lines in Fig.~\ref{fvpeaks}a (48\degr$\pm$6\degr). The other
parameters were varied over suitable ranges. The best fit was obtained
for \xs=$20^{\rm h}14^{\rm m}26\fs025$, \ys=41\degr13\arcmin32\farcs60,
\mbox{\Vs=--4.3~\kms,} $\psi$=59\degr, \Ms=12~\Msun, and $A$=0.42.

\begin{figure}
\centering
\resizebox{8.5cm}{!}{\includegraphics[angle=0]{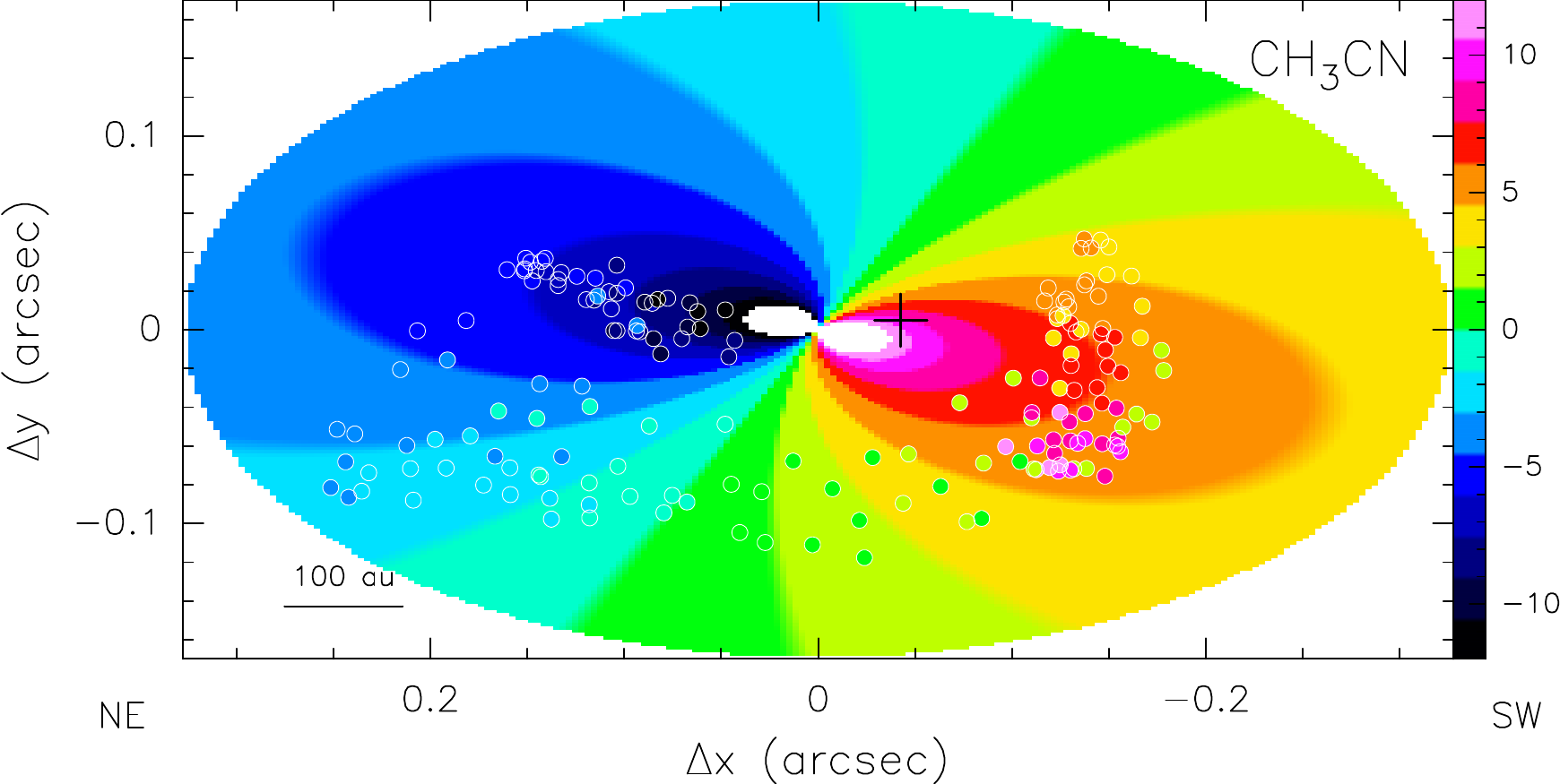}}
\caption{
Peaks of the \MCN(12--11) $K$=2,3,4,5,7,8 components in the different
velocity channels overlaid on the best-fit model of a disk undergoing
Keplerian rotation and radial infall (see text). The data have been rotated
counterclockwise by 42\degr\ to align the disk with the X-axis. The offsets
are relative to the disk center. The color indicates the velocity along
the line of sight relative to the the star. The black cross marks the
position of the 1.4~mm continuum peak.
}
\label{fkfm}
\end{figure}

A comparison between the data and the model is shown in Fig.~\ref{fkfm},
where the ellipse is the projection of the disk on the plane of the sky
and the circles are the observed \MCN\ peaks. The color indicates the
component of the velocity along the 
line of sight (LOS) and the agreement between model
and data is good where the color of the point matches that of the
disk model at the same position. Overall, the fit is satisfactory
and proves that a non-negligible radial velocity component (i.e., $A>0$)
is needed to reproduce the observed velocities. However, a significant
discrepancy between model and data is visible at the most red-shifted
velocities, suggesting that the SW part of the disk must be perturbed.
We discuss this issue in Sect.~\ref{sdisc}.

\begin{figure}
\centering
\resizebox{8.5cm}{!}{\includegraphics[angle=0]{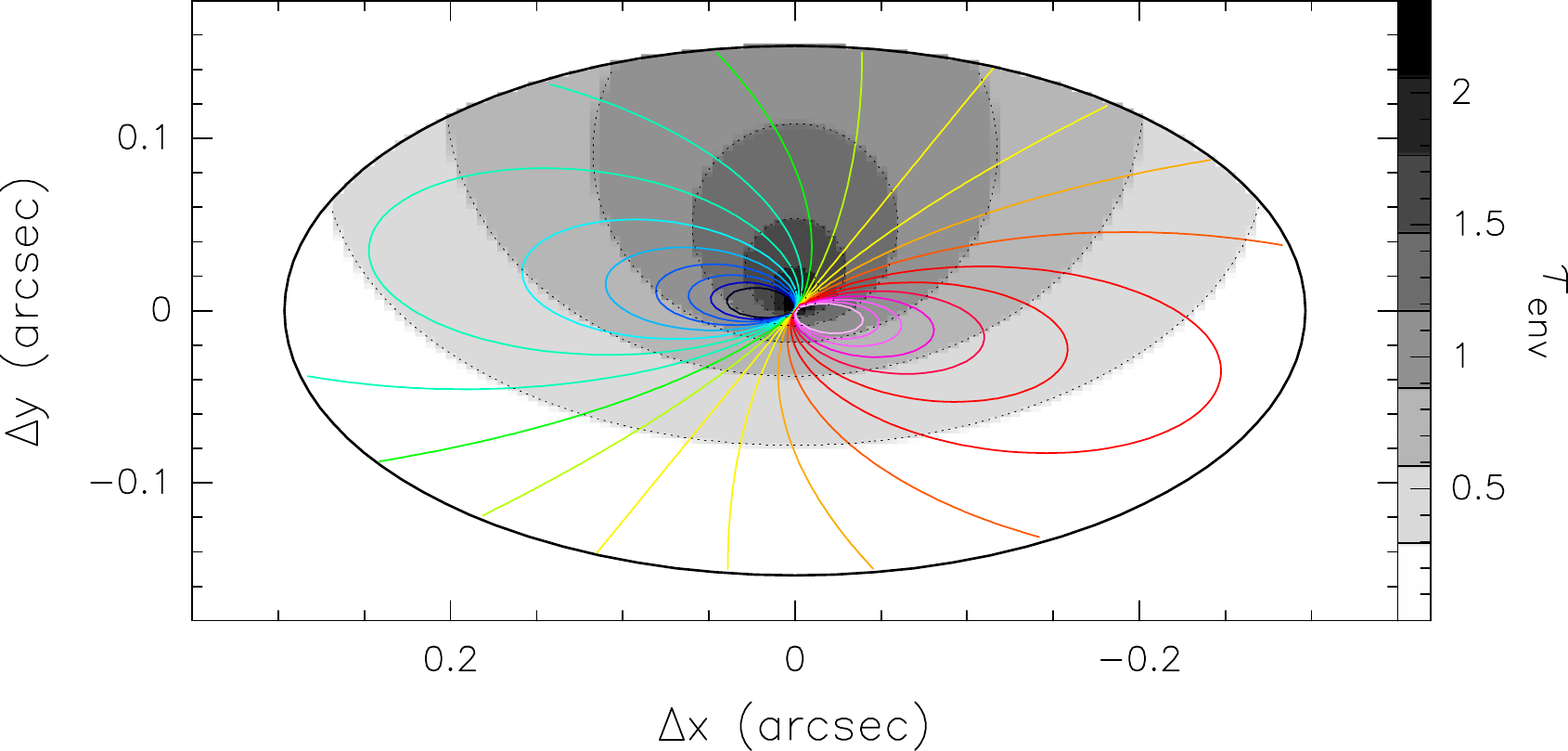}}
\caption{
Maps of the velocity and dust opacity over the disk according to our model.
The black ellipse outlines the border of the disk and the color-coded
patterns are the loci of points of the disk with the same velocity along
the line of sight, ranging from --10 to +10 in steps of 1~\kms. The gray scale with dotted contours is a map of the envelope dust opacity.
}
\label{ftvd}
\end{figure}

\paragraph{Fit of the C-shaped pattern}
~\\[3mm]
Now that we have a reasonable disk model, we can propose an explanation of
the C-shaped structures shown in Figs.~\ref{ffwhp} and~\ref{fvpeaks}. The
underlying hypothesis is that the emission from the part of the disk lying beyond
the plane of the sky passing through the disk center is more absorbed than
that lying closer to the observer. This hypothesis is supported by
Fig.~\ref{ftvd}, where we show the velocity patterns of the disk overlaid on
the map of the dust optical depth of the envelope between the disk and the
observer -- computed from Eq.~(\ref{eatau}). We note that the largest
opacities are found for positive values of $\Delta y$, which favors the
emission for $\Delta y<0$. Therefore, when we fit the line emission in each
velocity channel with a 2D Gaussian, the peak position tends to be skewed
towards negative $\Delta y$. This effect is important for the low-velocity
emission (green and yellow patterns in the figure) which is extended but
becomes progressively negligible at high velocities (red and blue patterns)
because the emission is more and more compact. This can explain why the blue-
and redshifted peaks in Fig.~\ref{ffwhp} lie close to the major axis of the
projected disk plane, whereas the lower-velocity peaks are shifted to the SE.

To prove that our explanation is viable, we used our model with
the best-fit parameters obtained above to generate a disk with surface
temperature computed from Eq.~(\ref{etrot}). We also fix the disk radius
to half the maximum size of the peak distribution in Fig.~\ref{fvpeaks},
namely $\Rd$=0\farcs3=490~au. As we assume the line emission to be
optically thick all over the disk, the observed
peak
line brightness temperature
at a given radius is equal to the kinetic temperature at this radius attenuated
by the absorption through the dusty envelope according to Eq.~(\ref{eatb}).
For our calculations, we assumed a Gaussian line profile with FWHM of 4~\kms.
To make the comparison with the observations more realistic, we also
convolved the disk+envelope model with the beam of our observations and
fitted the resulting cube channel by channel with a 2D Gaussian, as done
for the data. In this process, the only free parameter is the opacity
of the envelope, that is, the value of $\tau_0$ (see Appendix~\ref{atbr}),
which was varied until a reasonable match between the observed peak \Tb\
and that obtained from the model was achieved for $\tau_0=0.4$. We note
that the result holds for any line because we assume the line emission
from the disk to be optically thick.

\begin{figure}
\centering
\resizebox{8.5cm}{!}{\includegraphics[angle=0]{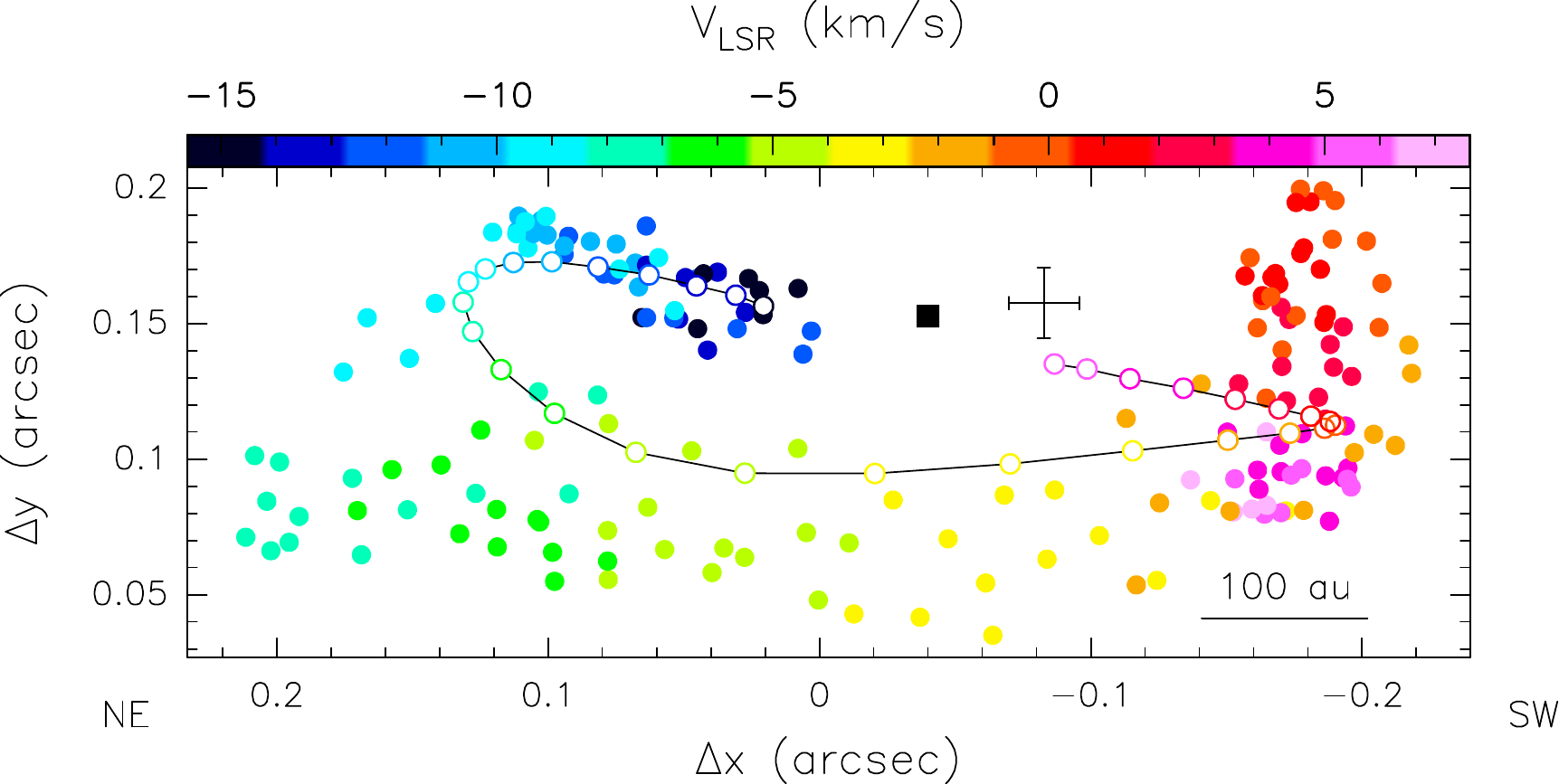}}
\caption{
Comparison between the peaks of the \MCN(12--11) $K$=2,3,4,5,7,8 components
in the different velocity channels (solid circles) and those obtained from
the model for a template line (empty circles connected by the solid line).
We note that the model peak distribution qualitatively reproduces the
C-shaped pattern described by the data.
The system is rotated clockwise by 42\degr\ and the offsets are relative
to the phase center of the observations.
The color indicates the LSR velocity. The black square marks the center
of the disk according to the model. The black cross indicates the position
of the 1.4~mm continuum peak with error bars.
}
\label{fmod}
\end{figure}

The distribution of the peaks obtained with our procedure is compared
in Fig.~\ref{fmod} to the peaks of the \MCN\ ground-state lines of
Fig.~\ref{fvpeaks}a. We conclude that the model qualitatively
mimics the C-shaped distribution of the data peaks, lending support to
our interpretation.

We believe that the proposed scenario could also explain the velocity drift
in Fig.~\ref{fpvmax}. As the part of the disk where the envelope opacity
is maximum is also the part of the disk where the infalling velocity is
blueshifted (see Fig.~\ref{ftvd}), this causes the line profile to become skewed,
whose blueshifted part is attenuated with respect to the redshifted
part. This effect must be more pronounced for the higher-energy transitions,
because they trace more internal regions, which are denser and as such more
affected by dust absorption. The result is that these lines appear more
redshifted than the lower-energy transitions.

Finally, it is worth computing the envelope mass. This is related to $\tau_0$
through the expression
\begin{equation}
M_{\rm env} = 2\pi \frac{\tau_0}{\kappa_{\rm d}} \Rd^2 \simeq 6.7~\Msun
   \label{emenv}
,\end{equation}
where $\kappa_{\rm d}$ is the dust
absorption coefficient that we assume to be equal to 0.01~cm$^2$g$^{-1}$.
A comparison between this value and the mass estimate obtained from the
continuum flux density measured by us ($\sim$280~mJy) is unreliable for
two reasons. First, the calculation requires knowledge of the halo dust
temperature. Second, a substantial fraction of the continuum emission is
resolved out by the interferometer, meaning that the mass estimated from the
continuum flux density is a lower limit. Vice versa, the mass computed
from Eq.~(\ref{emenv}) is not affected by this problem because $\tau_0$
takes into account the dust absorption due to the whole envelope being crossed
by the LOS.

\section{Discussion}
\label{sdisc}

The analysis performed in the previous sections provides us with some
hints as to the nature and evolutionary phase of \I.
First of all, we note that the proposed model to explain the C-shaped
structure of the peak distribution only holds if the disk is inclined in
such a way that the NW part of it lies beyond the star with respect to
the observer. It is only in this case that the opacity plotted in Fig.~\ref{ftvd} is maximum to the NW and the arc of the C-shaped structure points to
the SE, as observed. The orientation of the jet suggests that this is
indeed the disk inclination, because the blueshifted lobe is pointing
to the NW. Interestingly, the other two sources for which a similar
C-shaped distribution has been found --- G35.20--0.74~N (S\'anchez-Monge et
al.~\cite{sanch13}) and G11.92--0.61~MM1 (Bayandina et al.~\cite{baya24}) ---
do not satisfy the same requirement, because the arc points towards the blue
lobe of the associated jet. We speculate that this difference is
due to the lack of a thick envelope around these two sources, and that the
C-shape might be caused by self-absorption due to a large line opacity in
a geometrically thick disk. In this scenario, \I\ should be more embedded
and thus presumably less evolved than G35.20--0.74~N and G11.92--0.61~MM1.

If the source is indeed young and massive, the accretion rate should
also be high. This can be estimated from Eq.~(4) of Cesaroni et
al.~(\cite{cesa99}), which in our case takes the form
\begin{equation}
 \dot{M}_{\rm acc} = 2\pi \Rd \, \mu m_{\rm H} \frac{N_{\rm col}(\Rd)}{X_{\rm CH_3CN}} \cos\psi \, A\sqrt{\frac{G\Ms}{\Rd}}
,\end{equation}
where $\mu$=2.8 is the mean molecular weight, $m_{\rm H}$ the
mass of the hydrogen atom, $G$ the gravitational constant, $X_{\rm
CH_3CN}$ the \MCN\ abundance relative to \HM, $\cos\psi$ a correction
factor due to the inclination of the disk, and $\Ncol$ is given by
Eq.~(\ref{encol}). For $\Ms$=12~\Msun, $\Rd$=0\farcs3=490~au, $A$=0.42
(see Sect.~\ref{smodd}), and $X_{\rm CH_3CN}$=$10^{-9}$ , we obtain
$\dot{M}_{\rm acc}\simeq1.7\times10^{-3}~M_\odot$\,yr$^{-1}$, a value in
reasonable agreement with that obtained by Cesaroni et al.~(\cite{cesa99})
from the millimeter continuum emission and the spread in peak velocity
among different \MCN\ lines ($\sim9.8\times10^{-4}~M_\odot$\,yr$^{-1}$).

The expressions of $\Trot$ and $\Ncol$ in Eqs.~(\ref{etrot}) and~(\ref{encol})
can also be used to estimate the Toomre $Q$ parameter (Toomre~\cite{toomre})
and hence investigate the disk stability. In our case, we have
\begin{equation}
Q = \sqrt{\frac{\Ms}{G\,R^3}} \frac{\sigma_{\rm v}\,X_{\rm CH_3CN}}{\pi\,\mu m_{\rm H} \Ncol}
,\end{equation}
where $\sigma_{\rm v}$ is the velocity dispersion. A lower limit on $Q$
can be obtained assuming that $\sigma_{\rm v}$ contains only a thermal
contribution, hence $\sigma_{\rm v}=\sqrt{3k\Tk/(\mu m_{\rm H})}$, where \Tk\
is kinetic temperature. Since the \MCN\ molecule is likely thermalized at
the disk densities, we further assume \Tk=$\Trot$ and can thus calculate
the lower limit on $Q$ as a function of the radius $R$ in the disk:
\begin{equation}
 Q > 1.5 \, \left(\frac{R}{\Rd}\right)^{-0.36}
.\end{equation}
Since
by definition
$R<\Rd$, one has $Q>1$ at all radii, which implies that
the disk is stable.

It is also interesting to compute the angle between the axis of the disk
and that of the jet. Using the same notation as in the Appendices,
we can express the directions of the two axes by means of the
unit vectors
\mbox{$\hat{e}_{\rm j}=(\sin\PAj\cos\phi,\cos\PAj\cos\phi,-\sin\phi)$} for the
jet and \mbox{$\hat{e}_{\rm d}=(\sin\PAd\sin\psi,\cos\PAd\sin\psi,-\cos\psi)$}
for the disk, where $\PAj$=--60\degr, $\phi$=3\degr, $\PAd$=--42\degr,
and $\psi$=59\degr. The angle between the two axes is thus equal to
$\arccos(\hat{e}_{\rm j}\cdot\hat{e}_{\rm d})\simeq33\degr$. Interestingly,
this value is very similar to the opening angle of the precession
cone (37\degr) estimated by Cesaroni et al.~(\cite{cesa05}). This
result lends further support to the scenario proposed by Shepherd et
al.~(\cite{shep}), Cesaroni et al.~(\cite{cesa05}), and Caratti o Garatti et
al.~(\cite{caga08}), which invokes the existence of precession to explain
the S-shaped morphology of the jet from \I.

\begin{figure}
\centering
\resizebox{8.5cm}{!}{\includegraphics[angle=0]{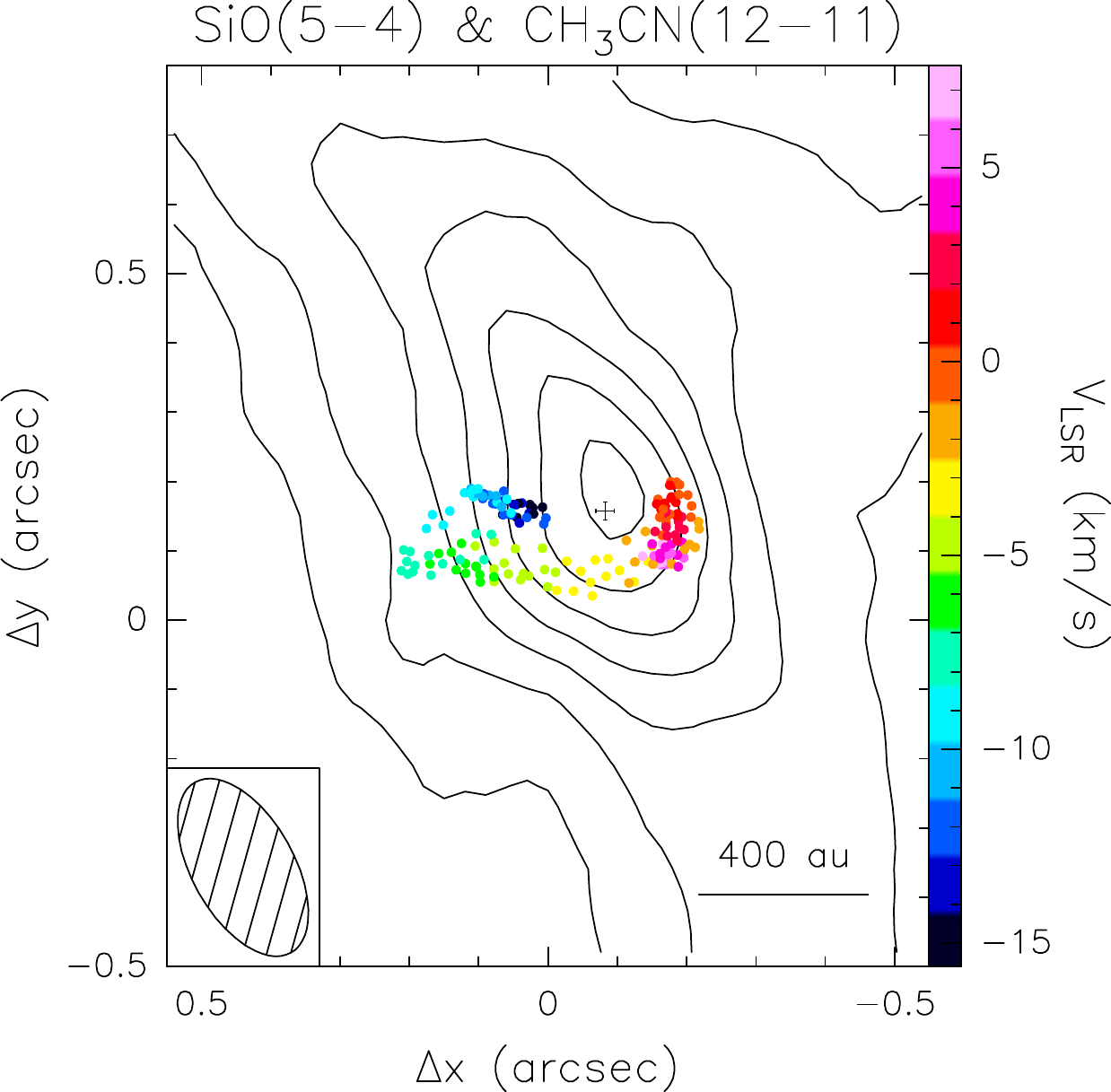}}
\caption{
Overlay of the \MCN\ peaks (circles) on the moment-8 map (contours) of
the SiO(5--4) line emission. The colored points and black cross are the
same as in Fig.~\ref{fmod}. The maps have been rotated by 42\degr\
counterclockwise for consistency with Fig.~\ref{fmod} and the offsets
are relative to the phase center of the observations. The contour levels
range from 6 to 36 in steps of 5~mJy/beam. The ellipse in the bottom left
corner denotes the synthesized beam of the SiO map.
}
\label{fsiomcn}
\end{figure}

As proposed by Shepherd et al.~(\cite{shep}), the precession could
be caused by tidal interactions with a lower-mass nearby (proto)star.
This hypothesis is consistent with the observed structure of the disk as
revealed by the peak distributions in Fig.~\ref{fvpeaks}. Independently
of the tracer, in this figure the distribution of the peaks is clearly
asymmetric with respect to the continuum peak and the points appear to
thicken on the SW side of it, whereas on the NE side they describe a more
extended pattern. This effect is even more clear in Fig.~\ref{fmod}, where
the largest discrepancy between the model and the observed peaks is seen
to the SW. A possible explanation for this asymmetry could be related to the
perturbing action of a nearby star lying to the SW of \I. The existence of
such a star is indeed suggested by the thermal jet detected at centimeter
wavelengths to the south of \I\ (Hofner et al.~\cite{hofn99,hofn07}) and is supported by the detection of a cluster of embedded YSOs in the area surrounding  \I\
 (Montes et al.~\cite{mont15}).

An alternative explanation for the perturbed structure of the redshifted
\MCN\ emission to the SW is that in this region the dense gas traced by
methyl cyanide is lifted from the disk plane by the jet. This possibility is
suggested by the comparison between the \MCN\ peaks and the SiO emission
(see Fig.~\ref{fsiomcn}). The latter appears to peak precisely in the gap
of the C-shaped \MCN\ pattern, where the continuum peak is located, and
the redshifted peaks are distributed along the SW side of the SiO jet.
Indeed, a tight interaction between the material rotating in the disk and
that expanding in the jet was also suggested by the LSR velocity and proper
motions of the \WAT\ masers (see Fig.~10 of Cesaroni et al.~\cite{cesa14}).
Therefore, we cannot exclude that a mechanism similar to that operating in
disk winds is affecting the structure of the circumstellar disk in \I.

\section{Summary and conclusions}
\label{ssum}

We report ALMA observations of the disk+jet system in the
massive (proto)star \I, with a substantial improvement on previous
observations at the same wavelengths with other instruments. The new data allow us to better define the geometrical and kinematical parameters
of the jet and to obtain maps of the temperature and column density in the
bow-shock regions at the tips of the jet lobes. As expected, both quantities
decrease moving away from the shocks in the
trailing
wake of the flow.
Our high-resolution images of the disk make it possible to derive the
rotation temperature and column density of the \MCN\ molecule as a function
of radius. We also show that the disk is not only undergoing Keplerian
rotation, but also has an inward velocity component implying an
accretion rate of $\sim$$10^{-3}$~\Msun~yr$^{-1}$, which is consistent with previous
estimates obtained by other means. With this knowledge of the physical structure
and velocity field of the disk, we were able to evaluate the Toomre $Q$
parameter, which indicates that stability is attained at all radii. Finally,
we propose that the disk morphology observed in high-density tracers might
be caused by a dusty envelope enshrouding the disk itself. Our analysis also
reveals the existence of a significant deviation from axial symmetry in the SW part of
the disk, which might be caused by a nearby, lower-mass companion
or could be the result of disk material being dragged into expansion along the jet.

\begin{acknowledgements}
V.M.R. acknowledges support from the grant PID2022-136814NB-I00 by the
Spanish Ministry of Science, Innovation and Universities/State Agency
of Research MICIU/AEI/10.13039/501100011033 and by ERDF, UE;  the grant
RYC2020-029387-I funded by MICIU/AEI/10.13039/501100011033 and by "ESF,
Investing in your future", and from the Consejo Superior de Investigaciones
Cient{\'i}ficas (CSIC) and the Centro de Astrobiolog{\'i}a (CAB) through
the project 20225AT015 (Proyectos intramurales especiales del CSIC); and
from the grant CNS2023-144464 funded by MICIU/AEI/10.13039/501100011033
and by “European Union NextGenerationEU/PRTR”.
A.S.-M. acknowledges support from the RyC2021-032892-I grant funded
by MCIN/AEI/10.13039/501100011033 and by the European Union `Next
GenerationEU’/PRTR, as well as the program Unidad de Excelencia María
de Maeztu CEX2020-001058-M, and support from the PID2020-117710GB-I00
(MCI-AEI-FEDER, UE).
This paper makes use of the following ALMA data: ADS/JAO.ALMA\#2016.1.00595.S
and ADS/JAO.ALMA\#2019.1.00199.S. ALMA is a partnership of ESO (representing
its member states), NSF (USA) and NINS (Japan), together with NRC (Canada),
NSC and ASIAA (Taiwan), and KASI (Republic of Korea), in cooperation with
the Republic of Chile. The Joint ALMA Observatory is operated by ESO,
AUI/NRAO and NAOJ.
\end{acknowledgements}

\begin{appendix}

\section{Jet model}
\label{appa}

\begin{figure}
\centering
\resizebox{8.5cm}{!}{\includegraphics[angle=0]{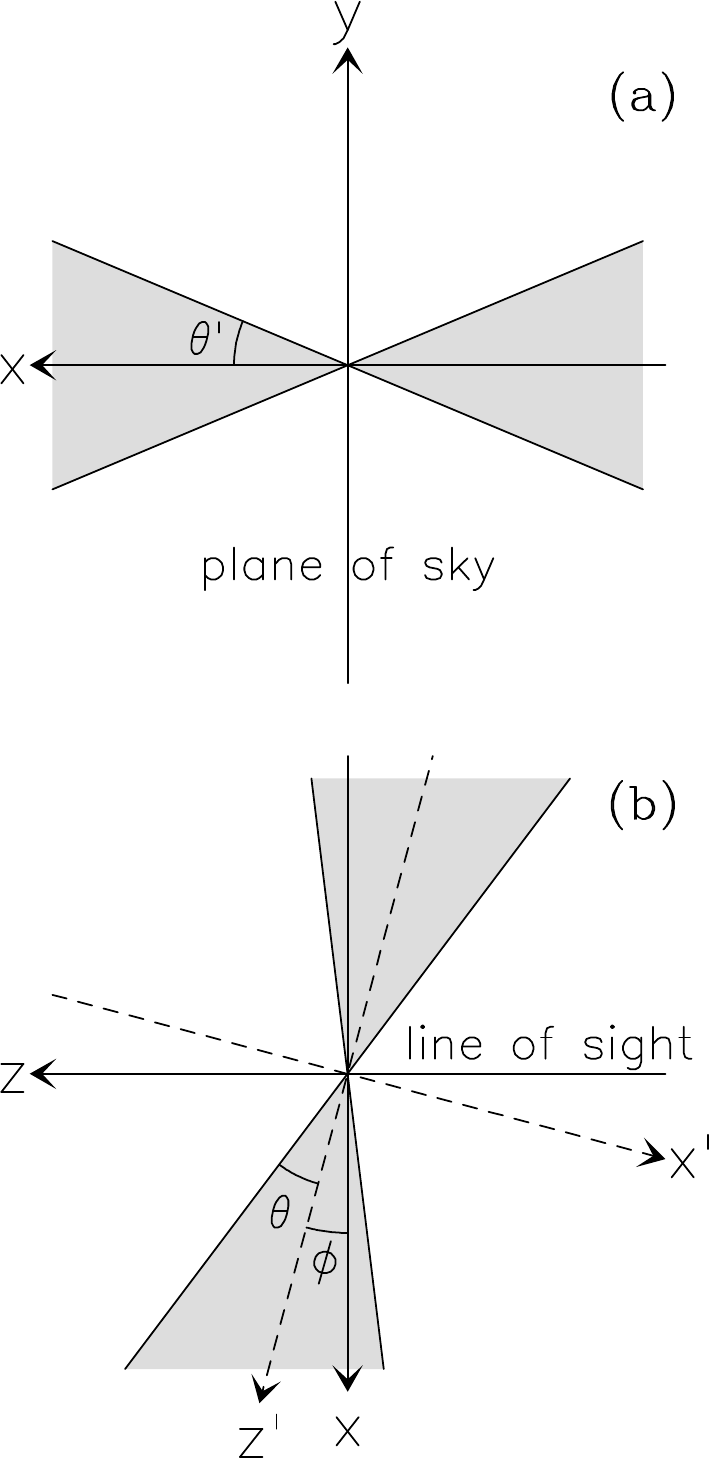}}
\caption{
Sketch of the proposed conical model for the jet.
{\bf a.} View of the jet in projection onto the plane of the sky. We indicate
the apparent opening angle with $\theta'$.
{\bf b.} Projection of the jet onto the $x,z$ plane. Here $\theta$ is the
true opening angle and $\phi$ the inclination angle of the jet axis with
respect to the plane of the sky. The observer lies at $z=-\infty$. The $z'$
axis coincides with the axis of the cone, while we choose $y'=y$.
}
\label{fajet}
\end{figure}

The jet model adopted by us is the same as the one presented by Cesaroni
et al.~(\cite{cesa99}). It consists of a cone with opening angle $\theta$
and axis inclined by $\phi$ on the plane of the sky. Figure~\ref{fajet}
shows a sketch, where we have chosen a coordinate system $(x,y,z)$ with $z$
parallel to the LOS and $x$ coincident with the projection of the cone
axis on the plane of the sky. The gas inside the cone is expanding with
radial velocity $V$ proportional to the distance $R$ from the vertex
(i.e., from the position of the star), according to the expression
\begin{equation}
 \vec{V} = V_0 \frac{\vec{R}}{R_0}
\end{equation}
where $V_0$ is the expansion speed at an arbitrary radius $R_0$. We stress
that our model is only sensitive to the ratio $V_0/R_0$.

We want to use the jet model to estimate the borders of the region inside
which emission is expected and compare it to the data, both in space and
velocity, to obtain the values of $\theta$, $\phi$, and $V_0/R_0$. On
the plane of the sky, such a region is represented by the gray area in
Fig.~\ref{fajet}a.
Comparison with the PV diagrams requires calculating the maximum/minimum
observed velocities attainable at a given position $x,y$.
The projection of the velocity along the LOS depends only on the
projection of vector $\vec{R}$, hence the observed LSR velocity is
\begin{equation}
 V_{\rm LSR} = V_{\rm sys}+V_{\rm z} = V_{\rm sys}+\frac{V_0}{R_0} z   \label{eavz}
\end{equation}
with $V_{\rm sys}=-3.9$~\kms\ systemic LSR velocity.
Consequently, at a given position $x,y$ the maximum/minimum values of
the observed velocity are attained for the maximum/minimum values of $z$ across
the cone, i.e., on the surface of the cone.
To calculate these values we consider the expression of the cone in the
coordinate system $(x',y',z')$ where $z'$ is the axis of the cone (see
Fig.~\ref{fajet}b):
\begin{equation}
 z'^2 \tan^2\theta = x'^2+y'^2.
\end{equation}
By applying the transformation of coordinates
\begin{eqnarray}
 x' & = & x \sin\phi - z \cos\phi \nonumber \\
 y' & = & y \\
 z' & = & z \cos\phi + z \sin\phi \nonumber
\end{eqnarray}
after some algebra one obtains the equation of the cone in the $(x,y,z)$
coordinate system
\begin{equation}
 z_\pm = \frac{x\sin\phi\cos\phi\pm\cos\theta\sqrt{F(x,y;\theta,\phi)}}{\cos^2\theta-\sin^2\phi}   \label{ezpm}
\end{equation}
where we have defined
\begin{equation}
 F \equiv x^2\sin^2\theta+y^2(\sin^2\phi-\cos^2\theta).
\end{equation}
The maximum/minimum velocities can be calculated by substituting
Eq.~(\ref{ezpm}) into Eq.~(\ref{eavz}).

The projection $\theta'$ of angle $\theta$ on the plane of the sky
(Fig.~\ref{fajet}a) is obtained for $F=0$ because in this case there must
be only one intersection between the LOS and the cone. This gives
\begin{equation}
 \theta' = \arctan\left(\frac{\sin\theta}{\sqrt{\cos^2\theta-\sin^2\phi}}\right).
\end{equation}

The previous equations allowed us to compute the border of the emitting
region both in the $x,y$ plane and in the $x,V_{\rm LSR}$ (for $y=0$) and
$y,V_{\rm LSR}$ (for fixed $x$) PV diagrams. This pattern was then compared
to the data by eye and the input parameters $\theta$, $\phi$, and $V_0/R_0$
were varied until a reasonable match with the observed emission was obtained
(see Fig.~\ref{fpvsio}). As initial guesses for this procedure we used the
values obtained by Cesaroni et al.~(\cite{cesa99}), namely $\phi$=9\degr,
$\theta$=21\degr, and $V_0/R_0$=8.3~\kms\,arcsec$^{-1}$. We also adopted
a PA of the jet axis $\PAj$=--60\degr\ and assumed that the position of
the star was that of the 1.4~mm continuum peak, i.e., $\alpha$(J2000)=$20^{\rm
h}14^{\rm m}26\fs022$ and $\delta$(J2000)=41\degr13\arcmin32\farcs58
(see Sect.~\ref{srdisk}). The best fit was obtained for $\theta$=10\degr,
$\phi$=3\degr, and $V_0/R_0$=10~\kms\,arcsec$^{-1}$.

\section{Disk+halo model}
\label{appb}

\begin{figure}
\centering
\resizebox{7.0cm}{!}{\includegraphics[angle=0]{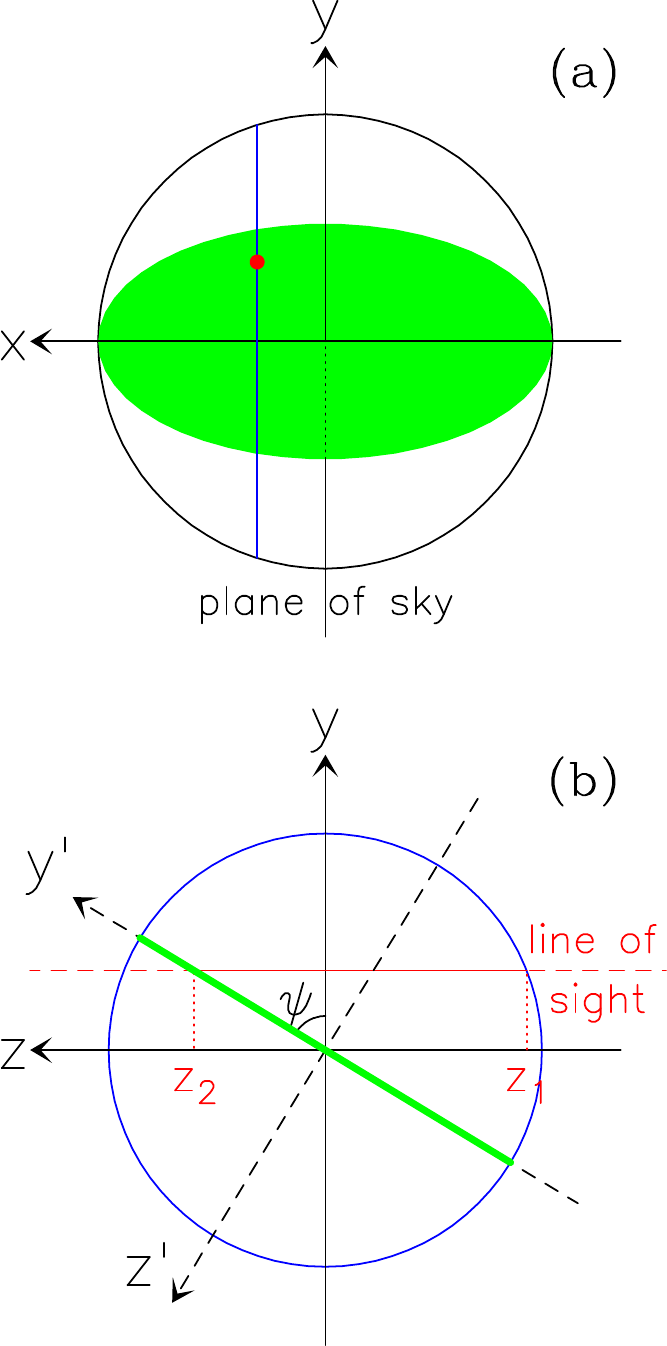}}
\caption{
Sketch of the proposed disk+halo model.
{\bf a.} View of the halo in projection onto the plane of the sky. The $y$
axis is the projection of the disk rotation axis. The circle is the
outer
border
of the spherical halo, while the green ellipse represents the disk. The
red point indicates a generic LOS and the blue line is the intersection
between the halo and the plane parallel to the disk axis and containing
the given LOS.
{\bf b.} Cut of the disk+halo corresponding to the blue line in panel a.
The green line denotes the disk seen edge on and the red line is the
LOS indicated by the red dot in panel a. Points $z_1$ and $z_2$
are the intersections with the halo. The $y'$ axis is rotated by the angle
$\psi$ in the $x,z$ plane with respect to
the plane of the sky. The observer lies at $z=-\infty$.
}
\label{fadisk}
\end{figure}

To explain the C-shaped pattern in Fig.~\ref{ffwhp}, we propose
a model consisting of a disk undergoing Keplerian rotation and radial
accretion
and a spherical halo extending from the center up to the outer radius, $\Rd$, of
the disk. Figure~\ref{fadisk} presents a sketch of the model,
where the disk lies on the plane $x',y'$, which forms an angle $\psi$ with the
plane of the sky $x,y$.

We assume the disk to be geometrically thin,
dust-free
and optically thick in the given transition, while the opacity of the halo
is due only to dust
with no line emission or absorption.
We also assume that the kinetic temperature of the
disk, \Tk, depends on the radius according to Eq.~(\ref{etrot}).

Our model is basically the same as the one used by S\'anchez-Monge et
al.~(\cite{sanch13}), with the addition of a radial velocity component
in the disk plus the spherical dusty envelope enshrouding it.
For the sake of comparison with the data, we need to calculate two physical
quantities: the observed LSR velocity, $V_{\rm LSR}$, and
line brightness temperature, $T_{\rm B}$.

\subsection{LSR velocity}
\label{avlsr}

The free parameters of the disk model are the position of the disk center,
i.e., of the star (\xs,\ys), the systemic velocity (\Vs), the position
angle of the projected major axis of the disk ($\PA$), the angle between
the disk plane and the plane of the sky ($\psi$), the mass of the star
(\Ms), and the ratio between the radial velocity and the rotation velocity
($A$). We assume that the accretion component of the velocity is $\propto
r^{-1/2}$
(with $r$ distance of a point of the disk from the center),
as in the case of free-fall. In practice, this means that $A$
does not depend on $r$ and ranges between 0, for pure Keplerian rotation,
and $\sqrt{2}$, when accretion proceeds like free-fall.

We wish to calculate the LSR velocity of the disk for a generic point $X,Y$
whose linear separation from the star (i.e., the disk center) is $X-\xs$
in right ascension and $Y-\ys$ in declination. We stress that in our
notation the offsets are in linear units (e.g., au).

The LSR velocity is given by the sum of the systemic velocity plus the
projection along the LOS of the rotation and infall velocities, therefore
one has
\begin{equation}
 V_{\rm LSR} = V_{\rm sys} + \sqrt{G\Ms} \frac{\pm x'-Ay'}{\left(x'^2+y'^2\right)^\frac{3}{4}} \sin\psi   \label{eavlsr}
\end{equation}
where the sign $\pm$ depends on the sense of rotation of the disk (in our case
the ``--'' sign was chosen). A point in the disk must satisfy the relationships
\begin{eqnarray}
x' & = & x \nonumber \\
y' & = & \frac{y}{\cos\psi} \label{eaxy} \\
z' & = & 0. \nonumber
\end{eqnarray}
where
\begin{eqnarray}
x & = & (X-\xs)\sin\PA + (Y-\ys)\cos\PA \label{eax} \\
y & = & -(X-\xs)\cos\PA + (Y-\ys)\sin\PA. \label{eay}
\end{eqnarray}

Through Eqs.~(\ref{eavlsr})--(\ref{eay}), one can finally express $V_{\rm LSR}$
as a function of $X$ and $Y$.

\subsection{Line brightness temperature}
\label{atbr}

The observed line brightness temperature of the
disk equals the kinetic temperature of the gas attenuated because
of the absorption by dust in the halo along the LOS, namely
\begin{equation}
 T_{\rm B} = T_{\rm K}(r) \exp\left[-\tau(x,y)\right]
  \label{eatb}
\end{equation}
with
\begin{equation}
 \tau(x,y) = \int_{z_1}^{z_2}\kappa_{\rm d} \, \rho \, \d{z}   \label{etau}
\end{equation}
where $r=\sqrt{x'^2+y'^2}$,
$\kappa_{\rm d}$ is the dust absorption coefficient,
$\rho$ the density of the halo, and $z_1$ and $z_2$ the
intersections
of the LOS with, respectively, the halo
and the disk (see Fig.~\ref{fadisk}b). For a generic LOS $x,y$, the latter
can be expressed as
\begin{eqnarray}
 z_1 & = & -\sqrt{\Rd^2-x^2-y^2} \\
 z_2 & = & y\tan\psi.
\end{eqnarray}
We note that $\sqrt{\Rd^2-x^2}$ is the radius of the circle that
corresponds to the intersection between the halo and the plane containing
the LOS and parallel to the disk axis (blue circle in Fig.~\ref{fadisk}b).

Under the assumption that the density of the halo scales like
$\rho=\rho_0(\Rd/R)$
with $R=\sqrt{x^2+y^2+z^2}$ distance of a point of the halo from the center,
Eq.~(\ref{etau}) can be written as
\begin{eqnarray}
 \tau(x,y) & = & \kappa_{\rm d}\,\rho_0\,\Rd \int_{z_1}^{z_2} R^{-1} \, \d{z} \nonumber \\
 & = & \tau_0
      \ln\left(\frac{y\tan\psi+\sqrt{x^2+y^2\left(1+\tan^2\psi\right)}}{\Rd-\sqrt{\Rd^2-x^2-y^2}}\right)  \label{eatau}
\end{eqnarray}
where $\tau_0\equiv\kappa_{\rm d}\,\rho_0\,\Rd$ is a dimensionless free parameter
of the model.

Finally, 
\Tk\ can be expressed as a function of $x$ and $y$ from Eqs.~(\ref{etrot}) and~(\ref{eaxy}):
\begin{equation}
 T_{\rm K}({\rm K})=158~\left[\frac{x^2+\left(\frac{y}{\cos\psi}\right)^2}{\Rd^2}\right]^{-0.16}.   \label{eatk}
\end{equation}
For our calculations, we have assumed a disk radius
$R_{\rm d}$=0\farcs3=490~au, estimated from the maximum size of the
distribution of the \MCN\ peaks in Fig.~\ref{fvpeaks}a.

In conclusion, the measured \Tb\ along a given LOS $x,y$ crossing the
disk can be computed by substituting Eqs.~(\ref{eatau}) and~(\ref{eatk})
into Eq.~(\ref{eatb}), while $x,y$ are related to the position in right
ascension and declination through Eqs.~(\ref{eax}) and~\ref{eay}).

\end{appendix}


\begin{thebibliography}{}

\bibitem[2020]{andr}
 Andrews, 2020, ARA\&A, 58, 483
\bibitem[2024]{baya24}
 Bayandina, O. S., Moscadelli, L., Cesaroni, R., et al. 2024, A\&A, submitted
\bibitem[2023]{beni}
 Benisty, M., Dominik, C., Follette, K., et al. 2023, in Protostars and
 Planets VII, ed. S. Inutsuka, Y. Aikawa, T. Muto, K. Tomida, and M. Tamura
 (Astronomical Society of the Pacific Conference Series), 605
\bibitem[2008]{caga08}
 Caratti o Garatti, A., Froebrich, D., Eisl\"offel, J., Giannini, T., \& Nisini, B. 2008, A\&A, 485, 137
\bibitem[1997]{cesa97}
 Cesaroni, R., Felli M., Testi L., Walmsley C.M., \& Olmi L. 1997, A\&A, 325, 725
\bibitem[1999]{cesa99}
 Cesaroni, R., Felli, M., Jenness, T., et al. 1999, A\&A, 345, 949
\bibitem[2005]{cesa05}
 Cesaroni, R., Neri, R., Olmi, L., et al. 2005, A\&A, 434, 1039
\bibitem[2011]{cesa11}
 Cesaroni, R., Beltr\'an, M.T., Zhang, Q., Beuther, H., \& Fallscheer, C. 2011, A\&A, 533, A73
\bibitem[2013]{cesa13}
 Cesaroni, R., Massi, F., Arcidiacono, C., et al. 2013, A\&A, 549, A146
\bibitem[2014]{cesa14}
 Cesaroni R., Galli D., Neri R., \& Walmsley C.~M. A\&A, 566, A73
\bibitem[2023]{cesa23}
 Cesaroni, R., Faustini, F., Galli, D., et al. 2023, A\&A, 671, A126
\bibitem[2016]{chen}
 Chen, H.-R. V., Keto, E., Zhang, Q., et al. 2016, ApJ, 823, 125
\bibitem[2018]{garu}
 Garufi, A., Benisty, M., Pinilla, P, et al. 2018, A\&A, 620, A94
\bibitem[2020]{vera}
 Hirota, T., Nagayama, T., Honma, M., et al. 2020, PASJ, 72, 50
\bibitem[1999]{hofn99}
 Hofner P., Cesaroni R., Rodr\'{\i}guez, L. F., \& Mart\'{\i}, J. 1999, A\&A, 345, L43
\bibitem[2007]{hofn07}
 Hofner P., Cesaroni R., Olmi L., et al. 2007, A\&A, 465, 197
\bibitem[2011]{johns}
 Johnston, K.G., Keto, E., Robitaille, T. P. Wood, K. 2011, MNRAS, 415, 2953
\bibitem[1999]{kawa99}
 Kawamura, J. H., Hunter, T. R., Tong, C.-Y. E., et al. 1999, PASP, 111, 1088
\bibitem[2010]{ketzha}
 Keto, E. \& Zhang, Q. 2010, MNRAS, 406, 102
\bibitem[2000]{ppiv}
 Kurtz, S., Cesaroni, R., Churchwell, E., Hofner, P., \& Walmsley, C.M. 2000, in
\bibitem[2023]{massi23}
 Massi, F., Caratti o Garatti, A., Cesaroni, R., et al. 2023, A\&A, 672, A113
\bibitem[2015]{mont15}
 Montes, V.A., Hofner, P., Anderson, C., \& Rosero, V. 2015, ApJS, 219, 41
\bibitem[2005]{mosca05}
 Moscadelli, L., Cesaroni, R., \& Rioja, M.J. 2005, A\&A, 438, 889
\bibitem[2011]{mosca11}
 Moscadelli, L., Cesaroni, R., Rioja, M.J., Dodson, R., \& Reid, M.J. 2011, A\&A 526, A66
\bibitem[2015]{palau15}
 Palau, A., Ballesteros-Paredes, J., V\'azquez-Semadeni, et al. 2015, MNRAS, 453, 3785
\bibitem[2008]{qiu08}
 Qiu, K., Zhang, Q., Megeath, S.Th., et al. 2008, ApJ, 685, 1005
\bibitem[2013]{sanch13}
 S\'anchez-Monge, \'A., Cesaroni, R., Beltr\'an, M.T., et al. 2013, A\&A, 552, L10
\bibitem[2018]{statcont}
 S\'anchez-Monge, \'A., Schilke, P., Ginzburg, A., Cesaroni, R., \& Schmiedeke, A. 2018, A\&A, 609, A101
\bibitem[2000]{shep}
 Shepherd, D.S., Yu, K.C., Bally, J., \& Testi, L. 2000, ApJ, 535, 833
\bibitem[2011]{su07}
 Su, Y.-N., Liu, S.-Y., Chen, H.-R., Zhang, Q., \& Cesaroni, R. 2007, ApJ, 671, 571
\bibitem[1964]{toomre}
 Toomre, A. 1964, ApJ, 139, 1217
\bibitem[1995]{walms}
 Walmsley, C. M. 1995, in Rev. Mexicana Astron. Astrofis. Ser. Conf. 1, Circumstellar Disks, Outflows, and Star Formation, ed. S. Lizano \& J. M. Torrelles (Mexico, DF: Inst. Astron., UNAM), 137
\bibitem[1990]{wilk90}
 Wilking, B. A., Blackwell, J. H., \& Mundy, L. G. 1990, AJ, 100, 758

\end{thebibliography}
\end{document}